\documentclass[10pt,letterpaper]{article}
\usepackage[top=0.85in,left=2.75in,footskip=0.75in]{geometry}

\usepackage{amsmath,amssymb}

\usepackage{changepage}

\usepackage{textcomp,marvosym}

\usepackage{cite}

\usepackage{nameref,hyperref}

\usepackage[right]{lineno}

\usepackage[nopatch=eqnum]{microtype}
\DisableLigatures[f]{encoding = *, family = * }

\usepackage[table]{xcolor}

\usepackage{array}

\newcolumntype{+}{!{\vrule width 2pt}}

\newlength\savedwidth

\newcommand\thickhline{\noalign{\global\savedwidth\arrayrulewidth\global\arrayrulewidth 2pt}%
\hline
\noalign{\global\arrayrulewidth\savedwidth}}

\raggedright
\usepackage[aboveskip=1pt,labelfont=bf,labelsep=period,justification=raggedright,singlelinecheck=off]{caption}

\makeatletter
\renewcommand{\@biblabel}[1]{\quad#1.}
\makeatother

\usepackage{lastpage,fancyhdr,graphicx}
\usepackage{epstopdf}
\fancyheadoffset[L]{2.25in}
\fancyfootoffset[L]{2.25in}
\usepackage[markup=underlined]{changes}
\definechangesauthor[color=violet]{ADW}
\definechangesauthor[color=magenta]{GD}
\definechangesauthor[color=red]{PS}

\reversemarginpar

\usepackage{siunitx}
\DeclareSIUnit{\molar}{M}

\newcommand{\ie}{\textit{i.e.}}

\newcommand{\unitstep}{H}

\begin{document}
\vspace*{0.2in}

\begin{flushleft}
{\Large
\textbf\newline{Inward rectifier potassium channels interact with calcium channels to promote robust and physiological bistability} 
}
\newline
De Worm Anaëlle\textsuperscript{1*},
Drion Guillaume\textsuperscript{1\Yinyang},
Sacré Pierre\textsuperscript{1\Yinyang}
\\
\bigskip
\textbf{1} Department of Electrical Engineering and Computer Science, University of Liège, Liège, Belgium
\\
\bigskip

%
%
\Yinyang These authors contributed equally to this work.





*anaelle.deworm@uliege.be

\end{flushleft}

\section*{Abstract}

Projection neurons in the dorsal horn relay nociceptive input to supraspinal centers. During central sensitization, a subset of them switches from tonic firing to plateau potentials with sustained afterdischarges, a change that requires intrinsic bistability between a resting and a spiking state. Voltage-gated L-type calcium (CaL) channels can produce bistability, but reach physiological resting states only when paired with voltage-gated potassium channels, most of which simultaneously shrink the bistability window. How robust, physiological bistability arises has therefore remained unclear.
Using a minimal conductance-based model, we show that inward rectifier potassium (Kir) channels enlarge the bistability window when combined with CaL channels, while M-type potassium (KM) channels slightly reduce it. Within the parameter region where bistability is both robust and physiological, both channel types can sustain bistability, but the CaL+Kir combination produces a substantially larger window and is more robust to noise and intrinsic variability.
This window-enlarging effect traces to a shape feature of the outward Kir steady-state current: like the CaL current, it has a region of negative differential conductance around the spike threshold, a feature absent from KM and from most other voltage-gated potassium currents. Bifurcation analysis further shows that the two pairs support qualitatively distinct excitability: plateau-generating bistability for CaL+Kir and resonator-like dynamics for CaL+KM. These conclusions hold in a two-compartment model of deep projection neurons with realistic ion channel complements, and identify the CaL+Kir pair as a candidate intrinsic mechanism for central sensitization.

\section*{Author summary}

Many neurons can hold two stable states (resting or firing) at the same input current, depending on their recent history. This bistability underlies a form of short-term memory at the single-cell level and appears in contexts as varied as sleep rhythms, motor control, working memory, and pain signal amplification during central sensitization.
How bistability arises in a physiologically plausible way has remained partly unresolved. L-type calcium (CaL) channels generate bistability, but at abnormally low resting states; additional channels must redress these potentials, yet most potassium channels that do so also shrink the range of currents over which bistability exists. 
Using a conductance-based model, we show that inward rectifier potassium (Kir) channels are more favourable than M-type potassium (KM) channels in this regard: paired with CaL channels, they produce a bistability that is both physiological and stable against noise and cell-to-cell variability over a substantially wider range of parameters than KM channels can. The difference traces to a shape feature of the Kir steady-state current, shared with CaL but absent from most potassium currents. Because Kir channels are regulated by many neuromodulatory pathways in the dorsal horn, the CaL+Kir pair is a candidate mechanism for pain amplification during central sensitization and, more broadly, for single-cell short-term memory.


\section*{Introduction}

Nociceptive pain protects the body by signaling tissue-damaging stimuli~\cite{woolf_what_2010}. Under strong or repeated stimulation, sensitization alters the nociceptive system and amplifies pain signaling~\cite{latremoliere_central_2009}. In the dorsal horn, projection neurons relay sensory input to supraspinal centers, and sensitization renders a subset of them hyperexcitable~\cite{reali_integrated_2005,monteiro_switching-and_2006,zain_alterations_2019}. Deep projection neurons, in particular, can switch from tonic firing to plateau potentials---a sustained depolarized state interposed between the spiking and resting voltage ranges---with sustained afterdischarges when neuromodulation changes~\cite{derjean_dynamic_2003}. These afterdischarges substantially increase the signal sent upstream and underlie hyperalgesia~\cite{cata_altered_2006,monteiro_switching-and_2006,robinson_altered_2014}. Because they sustain spiking beyond the end of a stimulus instead of reverting to rest, sustained afterdischarges reveal an intrinsic bistability between resting and spiking states~\cite{monteiro_switching-and_2006}.

Neuronal bistability has been observed in many populations~\cite{lee_bistability_1998,kazantsev_bistability_2011,dovzhenok_exploring_2012,engbers_bistability_2013,franci_balance_2013,borges_roles_2023}, yet how it emerges in single neurons in a form that is both robust and physiological remains unclear. We call bistability \textit{physiological} when its resting states lie above the potassium reversal potential, as observed in deep projection neurons~\cite{morisset_ionic_1999,derjean_dynamic_2003,reali_intrinsic_2011}; we call it \textit{robust} when the \textit{bistability window}, defined as the range of currents from $I_1$ (onset of spiking from rest) to $I_2$ (cessation of resting from spiking), is wide enough that both states are reliably reached. Computational studies show that voltage-gated calcium channels create bistability, but at resting states that become physiological only when these channels are paired with voltage-gated potassium channels~\cite{crunelli_window_2005,le_bistable_2006,borges_roles_2023}. Most such potassium channels, however, also shrink the bistability window~\cite{yuen_bistability_1995,franci_balance_2013,naudin_general_2023}, producing a trade-off between window size and the plausibility of resting states. Inward rectifier potassium (Kir) channels, a distinct subclass expressed in superficial and deep dorsal horn projection neurons~\cite{derjean_dynamic_2003,murata_neuronal_2016,ford_inward-rectifying_2016,malcangio_gabab_2018,brewer_enhanced_2018}, can themselves induce bistability~\cite{shoemaker_neural_2011,sanders_nmda_2013,amarillo_inward_2018,delmoe_conditions_2023}. Whether and how Kir channels interact with calcium channels to shape bistability, and through what mechanism, remain open questions.

Here, we use a \emph{minimal} conductance-based model to show that Kir channels, when coupled with L-type calcium (CaL) channels, enlarge the bistability window, whereas M-type potassium (KM) channels reduce it. We reproduce this effect in a published two-compartment model of deep projection neurons that already includes CaL and Kir channels alongside other currents~\cite{le_franc_multiple_2010}. The two pairs produce bistabilities that differ in robustness to noise and intrinsic variability; we trace these differences to the contrasting shapes of the Kir and KM steady-state currents, and relate each pair to a distinct excitability switch.

\section*{Results}

We begin by asking how each channel pair shapes the bistability window in a minimal model, then validate the effect in a two-compartment model. 
We then quantify the robustness of each bistability to noise and intrinsic variability, tracing the differences to the steady-state shapes of the three currents.
Finally, we link each pair to a distinct excitability switch, which unifies the mechanistic and functional observations.

\subsection*{CaL channels drive bistability; KM channels slightly narrow the window while Kir channels enlarge it}

To examine how calcium and potassium channels jointly shape bistability, we built a minimal conductance-based model combining a fast sodium current $I_\mathrm{Na}$, a delayed-rectifier potassium current $I_\mathrm{KDR}$, a leak current $I_\mathrm{leak}$, and L-type calcium (CaL) channels, and compared the effect of adding either M-type potassium (KM) or inward rectifier potassium (Kir) channels.

\begin{figure}[h!]
    \begin{center}
        \includegraphics[width=\textwidth]{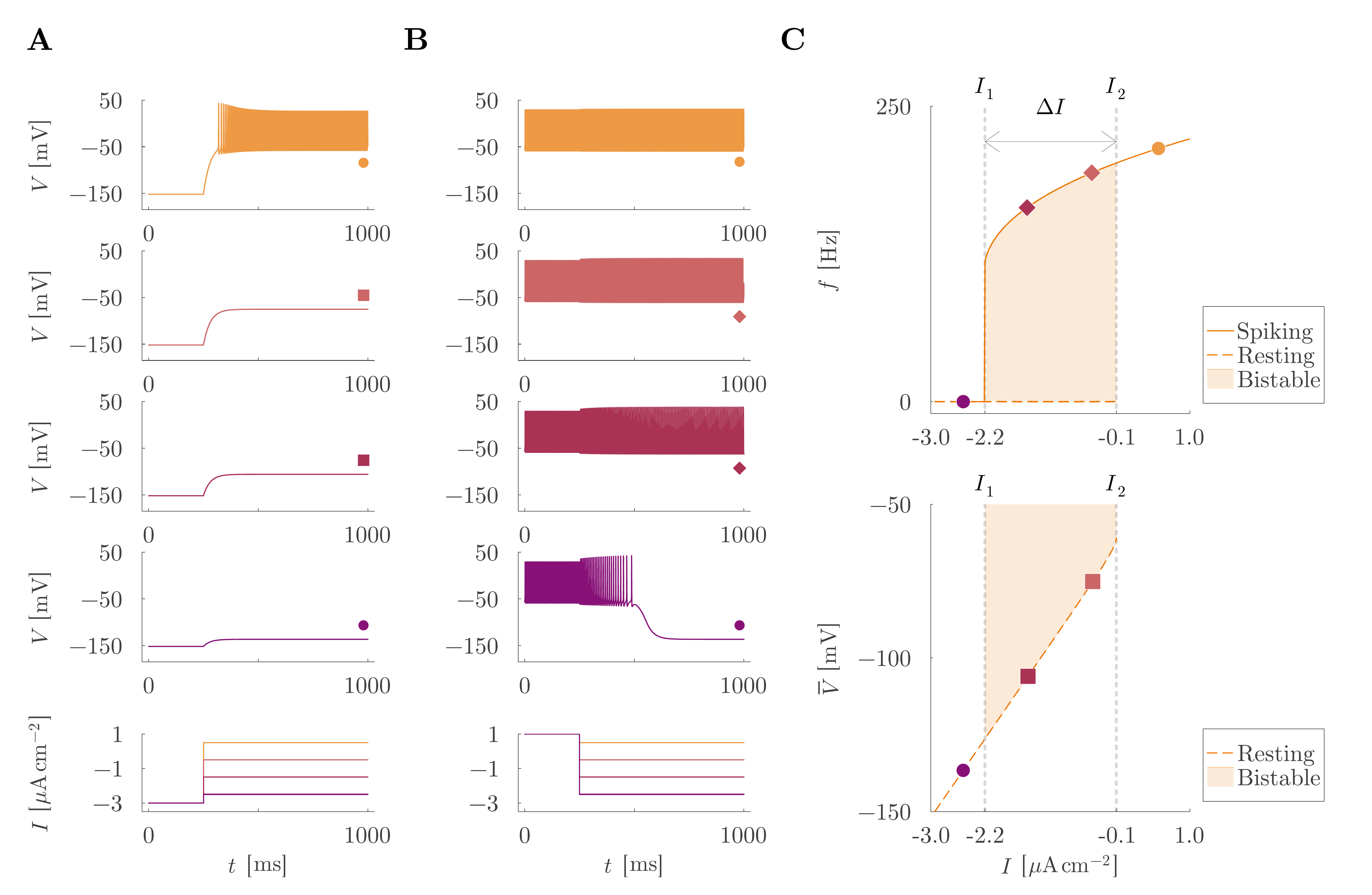}
        \caption{{
        Voltage-gated calcium channels produce bistability between resting and spiking states.
        A:~From an initial resting state, ascending current steps progressively depolarize the membrane (purple and reds). The largest step triggers a switch to spiking (yellow).
        B:~From an initial spiking state, descending current steps progressively decrease the firing frequency (yellow and reds). The largest step triggers a switch to resting (purple).
        C:~The frequency-current (fI) curves obtained from initially resting and initially spiking neurons (top, dashed and solid lines, respectively) identify the bistability window: the current range from $I_1$ to $I_2$ in which both states coexist (red tones, matching A and B). The resting equilibria, noted $\overline{V}(I)$, depolarize progressively with increasing current (bottom). Markers in A, B, and C identify the final state observed for each of the final applied currents of $-2.5$, $-1.5$, $-0.5$, and $\qty{0.5}{\micro\ampere\per\centi\meter^2}$ (from bottom to top).
        {\label{fig:1}} 
        }}
    \end{center}
\end{figure}

CaL channels alone created a bistability window, but at resting states that fell below the potassium reversal potential. Without CaL channels the minimal model did not exhibit bistability (see~\nameref{app:fI_hh}). Adding CaL channels produced bistability between a resting and a spiking state, revealed by applying ascending and descending current steps that reached the same final current (Fig.~\ref{fig:1}A,B). Indeed, for the two middle currents, the final state depended on the initial behavior (Fig.~\ref{fig:1}A,B, red tones), while the smallest and largest currents yielded resting or spiking regardless of initial conditions (purple and yellow). 
The frequency-current (fI) curves obtained from initially resting and
initially spiking neurons (see Methods) showed a bistability window of
$\qty{2.1}{\micro\ampere\per\centi\meter^2}$ delimited by $I_1$ and $I_2$
(Fig.~\ref{fig:1}C, top). Above $I_2$, only spiking was observed (Fig.~\ref{fig:1}, yellow marker). From $I_1$ to $I_2$, both resting and spiking could be observed depending on the neuron initial conditions (Fig.~\ref{fig:1}, markers in red tones). Only resting was observed for currents below $I_1$ (Fig.~\ref{fig:1}, purple marker). 
Throughout, we use $\overline{V}(I)$ to denote
the steady-state membrane potential associated with the resting equilibrium
at a given applied current $I$; this is distinct from the conventional
resting potential, which refers to the membrane potential in the absence of
stimulation. Within this window, $\overline{V}(I)$ fell well below the
potassium reversal potential (Fig.~\ref{fig:1}C, bottom), so CaL channels
must be combined with other currents to produce physiologically plausible
bistability.

\begin{figure}[h!]
    \begin{center}
        \includegraphics[width=\textwidth]{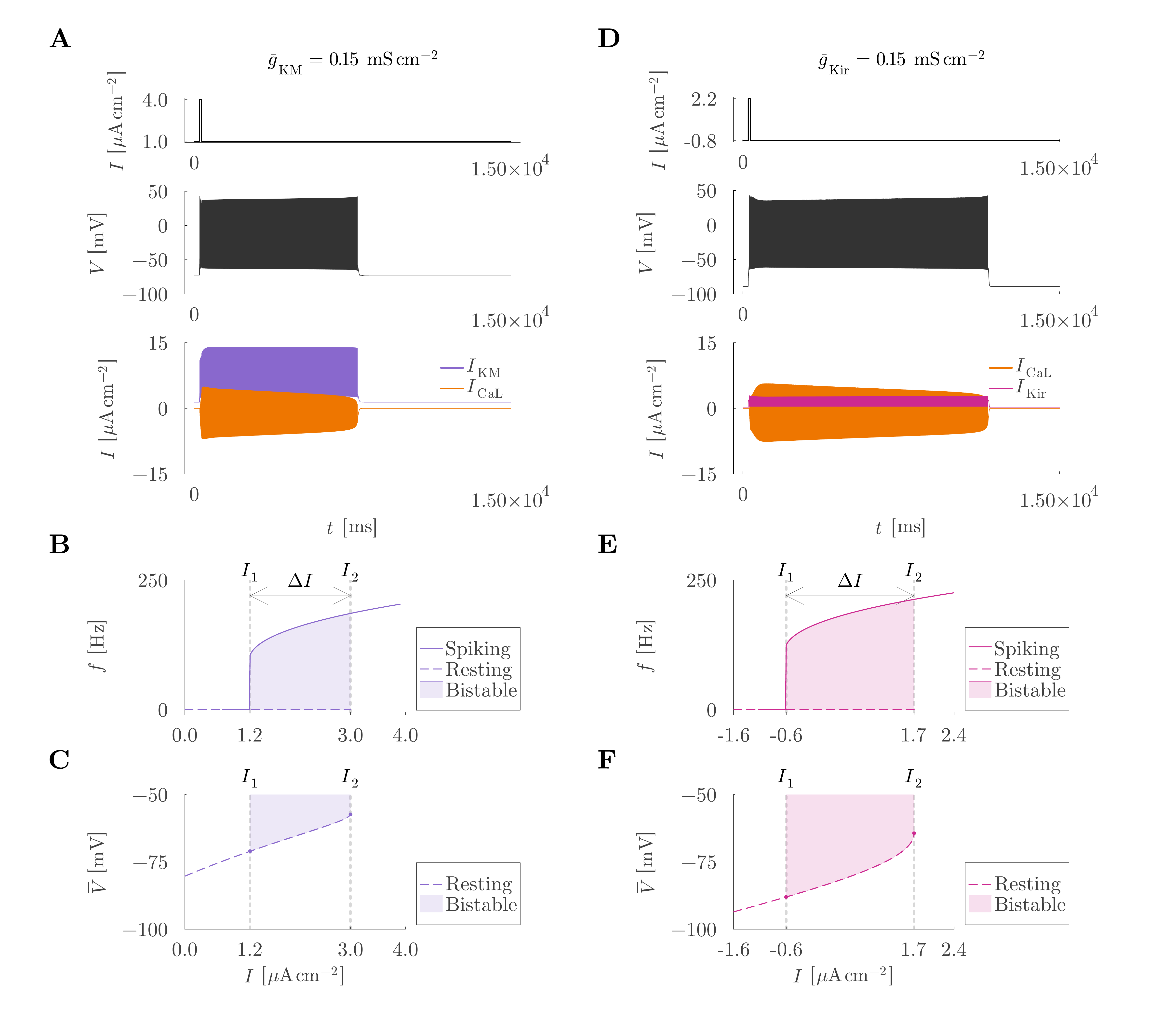} 
        \caption{{%
        Both KM and Kir channels redress the resting equilibria, and modulate bistability.
        A:~After adding an M-type potassium conductance $\bar{g}_\mathrm{KM}$, a current pulse from a baseline below $I_1$ elicits a finite afterdischarge: the neuron spikes during the pulse and returns to rest afterwards (top). The  $I_\mathrm{CaL}$ and $I_\mathrm{KM}$ traces show that $I_\mathrm{CaL}$ declines during spiking, eventually terminating it (bottom).
        B:~The fI curves show that the bistability window created by CaL channels (Fig.~\ref{fig:1}C, top) slightly narrows after adding KM channels.
        C:~The resting equilibria within the window, originally below $E_K$ (Fig.~\ref{fig:1}C, bottom), are redressed by KM channels.
        D:~After replacing KM with an inward rectifier potassium conductance $\bar{g}_\mathrm{Kir}$, a current pulse from a baseline below the new $I_1$ again elicits a finite afterdischarge.
        E:~The bistability window is preserved after adding Kir channels.
        F:~Kir channels also redress the resting equilibria within the window.
        {\label{fig:2}}
        }}
    \end{center}
\end{figure}

Adding an M-type potassium maximal conductance $\bar{g}_\mathrm{KM}$ redressed the resting equilibria but slightly narrowed the bistability window. With a baseline current below the new $I_1$, a current pulse elicited a finite afterdischarge: the neuron transitioned to spiking during the pulse and eventually returned to resting after the pulse ended (Fig.~\ref{fig:2}A). The termination of spiking is attributable to the progressive decrease in the CaL current caused by the slow cumulative inactivation of CaL channels during spiking (Fig.~\ref{fig:2}A, bottom). The corresponding fI curves showed a narrower window than without KM channels (Fig.~\ref{fig:2}B, purple, vs.\ Fig.~\ref{fig:1}C), while $\overline{V}(I)$ within the window approached physiological values (Fig.~\ref{fig:2}C). 
The stimulation protocols in Fig.~\ref{fig:2}A and Fig.~\ref{fig:1}A–B are equivalent approaches that can demonstrate bistability. In the pulse protocol, the pulse amplitude must exceed the bistability window to trigger a transition from resting to spiking, while the baseline current must lie within it to sustain the new state. In the ascending/descending step protocol, the pre-step current must fall above $I_2$ or below $I_1$ to establish the initial state, while the final current must lie within it to maintain that state.

Replacing KM with an inward rectifier potassium (Kir) conductance $\bar{g}_\mathrm{Kir}$ preserved the correction of resting equilibria while enlarging the bistability window. A current pulse of the same amplitude and a baseline below $I_1$ again produced a finite afterdischarge, with the neuron returning to a resting equilibrium of $\qty{-89}{\milli\volt}$ (Fig.~\ref{fig:2}D). As with KM channels, spiking termination is attributable to the slow cumulative inactivation of CaL channels (Fig.~\ref{fig:2}D, bottom). 
The superimposed fI curves showed a larger bistability window than with CaL alone, shifted toward higher currents (Fig.~\ref{fig:2}E). Kir channels also brought $\overline{V}(I)$ within the window to physiological values (Fig.~\ref{fig:2}F).

\begin{figure}[t!]
    \begin{center}
        \includegraphics[width=\textwidth]{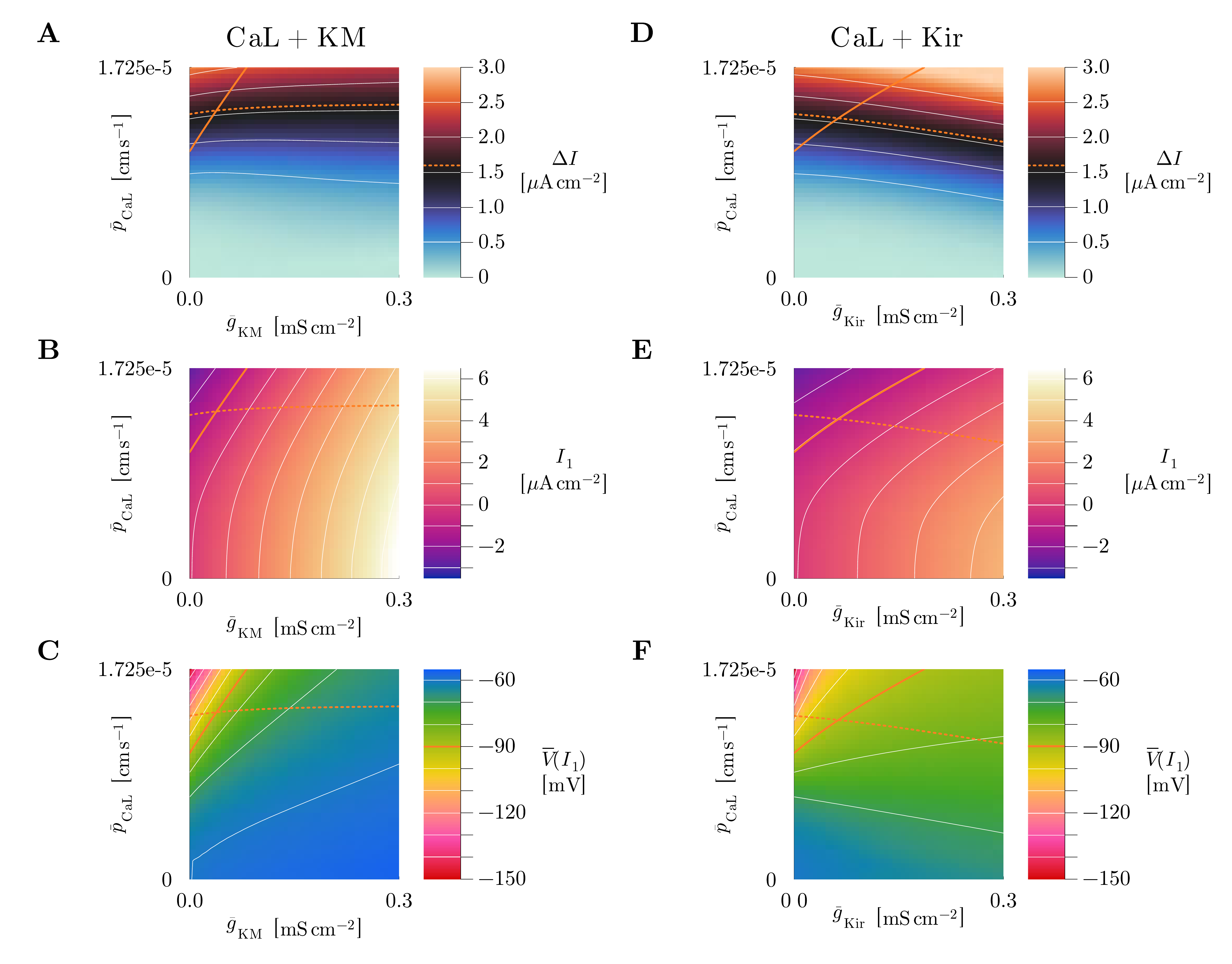} 
        \caption{{%
        KM and Kir channels have distinct effects on the bistability window size, while both redress the resting equilibria.
        A--C:~Heatmaps of the bistability window size $\Delta I = I_2 - I_1$ (A), its lower limit $I_1$ (B), and the corresponding resting equilibrium $\overline{V}(I_1)$ (C), as a function of $\bar{p}_\mathrm{CaL}$ and $\bar{g}_\mathrm{KM}$.
        D--F:~Same quantities as a function of $\bar{p}_\mathrm{CaL}$ and $\bar{g}_\mathrm{Kir}$. Each heatmap contains $51 \times 47$ data points ($\bar{g}_\mathrm{KM}$ and $\bar{g}_\mathrm{Kir}$ in steps of $\qty{0.006}{\milli\siemens\per\centi\meter^2}$; $\bar{p}_\mathrm{CaL}$ in steps of $\qty{3.75e-7}{\centi\meter\per\second}$). White lines are contour lines. The red solid line marks $\overline{V}(I_1)=\qty{-90}{\milli\volt}$ and the red dashed line marks $\Delta I = \qty{1.6}{\micro\ampere\per\centi\meter^2}$. 
        The two red lines together delimit the parameter region (top right) in which bistability is simultaneously robust ($\Delta I \geq \qty{1.6}{\micro\ampere\per\centi\meter^2}$) and physiological ($\overline{V}(I_1) \geq \qty{-90}{\milli\volt}$).
        {\label{fig:3}}       }}
    \end{center}
\end{figure}

To see how these effects evolve across the full conductance space, we swept $\bar{p}_\mathrm{CaL}$ jointly with either $\bar{g}_\mathrm{KM}$ or $\bar{g}_\mathrm{Kir}$. A systematic sweep over $\bar{p}_\mathrm{CaL}$ and $\bar{g}_\mathrm{KM}$ confirmed the observations above. We computed the bistability window size $\Delta I = I_2 - I_1$, its lower limit $I_1$, and the resting equilibrium at that limit $\overline{V}(I_1)$ across the two-dimensional parameter space (Fig.~\ref{fig:3}A--C, respectively). KM channels slightly narrowed the window at high $\bar{p}_\mathrm{CaL}$: with $\bar{p}_\mathrm{CaL}=\qty{1.725e-5}{\centi\meter\per\second}$, raising $\bar{g}_\mathrm{KM}$ to $\qty{0.3}{\milli\siemens\per\centi\meter^2}$ reduced $\Delta I$ from $\qty{2.7}{}$ to $\qty{2.3}{\micro\ampere\per\centi\meter^2}$ (Fig.~\ref{fig:3}A). For lower $\bar{p}_\mathrm{CaL}$, KM channels slightly enlarged $\Delta I$, but only in parameter regions where bistability was neither robust ($\Delta I < \qty{1.6}{\micro\ampere\per\centi\meter^2}$) nor physiological ($\overline{V}(I_1) < \qty{-90}{\milli\volt}$). KM channels raised both $I_1$ (Fig.~\ref{fig:3}B) and $\overline{V}(I_1)$ steeply, with $\overline{V}(I_1)$ reaching values above $\qty{-60}{\milli\volt}$ (Fig.~\ref{fig:3}C, blue area); this depolarization already hints at an excitability change driven by KM recruitment, which we examine later.

In contrast to this small narrowing effect, a sweep over $\bar{p}_\mathrm{CaL}$ and $\bar{g}_\mathrm{Kir}$ showed that Kir channels enlarge the bistability window while anchoring the resting equilibria. For any $\bar{p}_\mathrm{CaL}$, raising $\bar{g}_\mathrm{Kir}$ increased both $\Delta I$ and $I_1$, and held $\overline{V}(I_1)$ close to $\qty{-75}{\milli\volt}$ (Fig.~\ref{fig:3}D--F, respectively). These effects of CaL, KM, and Kir channels on window size were independent of the specific CaL current model used (see \nameref{app:other_CaL}). The contrasting behaviors of the CaL+KM and CaL+Kir pairs suggest that each relies on a distinct mechanism and may drive a distinct excitability switch, which we return to later.

\begin{figure}[h!]
    \begin{center}
        \includegraphics[width=\textwidth]{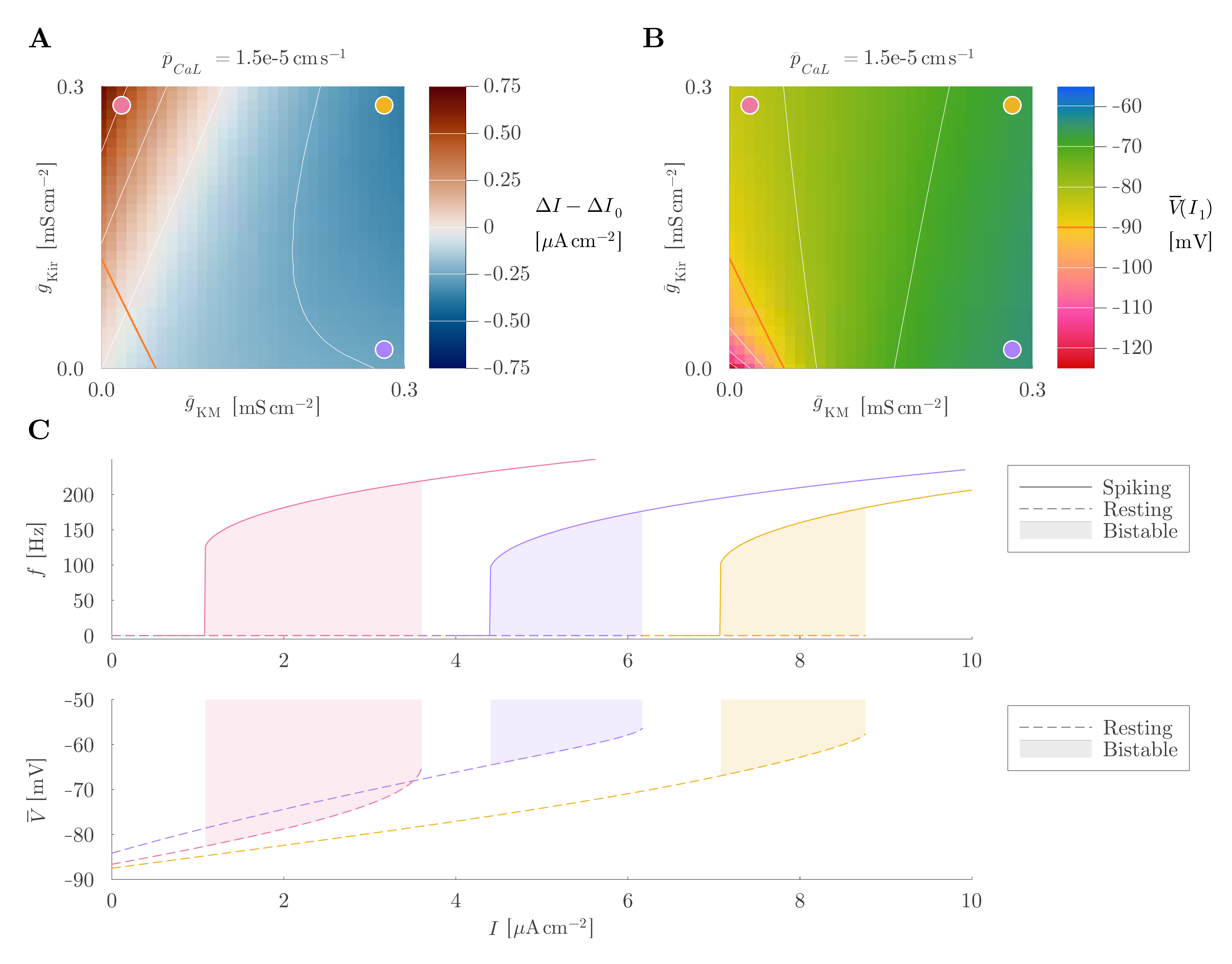} 
        \caption{{%
        When combined, Kir and KM channels retain their individual effects on the bistability window.
        A:~Heatmap of the change in bistability window size, $\Delta I - \Delta I_0$, as a function of $\bar{g}_\mathrm{KM}$ and $\bar{g}_\mathrm{Kir}$, with $\bar{p}_\mathrm{CaL}=\qty{1.5e-5}{\centi\meter\per\second}$ fixed. $\Delta I_0 = \qty{2.03}{\micro\ampere\per\centi\meter^2}$ denotes the window size at $\bar{g}_\mathrm{KM}=\bar{g}_\mathrm{Kir}=0$.
        B:~Heatmap of $\overline{V}(I_1)$ over the same parameter grid.
        C:~fI curves (top) and corresponding resting equilibria (bottom) for the three parameter sets marked in A and B. Each heatmap contains $31 \times 31$ data points. White lines are contour lines. The red solid line in A and B marks $\overline{V}(I_1)=\qty{-90}{\milli\volt}$.
        {\label{fig:altgt}} }}
    \end{center}
\end{figure}

A natural question is whether these effects are still observed when KM and Kir are expressed simultaneously. When KM and Kir channels were combined with CaL channels together, each retained its distinct effect.
Starting from a baseline bistability window of $\Delta I_0 = \qty{2.03}{\micro\ampere\per\centi\meter^2}$ (with $\bar{p}_\mathrm{CaL}=\qty{1.5e-5}{\centi\meter\per\second}$), we examined how co-varying $\bar{g}_\mathrm{KM}$ and $\bar{g}_\mathrm{Kir}$ (each from $0$ to $\qty{0.3}{\milli\siemens\per\centi\meter^2}$) affected $\Delta I$ (Fig.~\ref{fig:altgt}A).
Consistent with Fig.~\ref{fig:3}A, increasing $\bar{g}_\mathrm{KM}$ reduced $\Delta I$. 
For $\bar{g}_\mathrm{KM}\leq\qty{0.15}{\milli\siemens\per\centi\meter^2}$, increasing $\bar{g}_\mathrm{Kir}$ monotonically increased $\Delta I$, while for higher $\bar{g}_\mathrm{KM}$ values, a small reduction in $\Delta I$ (up to $\qty{0.11}{\micro\ampere\per\centi\meter^2}$) was observed as $\bar{g}_\mathrm{Kir}$ increased. This local reduction likely reflects a transient phenomenon originating from higher-order interactions between the two conductances, as a similar drop (at most $\qty{0.05}{\micro\ampere\per\centi\meter^2}$ was observed for $\bar{g}_\mathrm{KM}$ between $0.15$ and $\qty{0.21}{\milli\siemens\per\centi\meter^2}$ at low $\bar{g}_\mathrm{Kir}$, followed by a continuous increase in $\Delta I$ as $\bar{g}_\mathrm{Kir}$ increased. Thus, while the effect of $\bar{g}_\mathrm{Kir}$ on $\Delta I$ may become non-monotonic when combined with high $\bar{g}_\mathrm{KM}$, it generally acts to increase $\Delta I$.  
$\overline{V}(I_1)$ rose more steeply with $\bar{g}_\mathrm{KM}$ than with $\bar{g}_\mathrm{Kir}$ (Fig.~\ref{fig:altgt}B), confirming that Kir channels maintain resting equilibria in a lower voltage range than KM channels, even when combined. This is further illustrated in Fig.~\ref{fig:altgt}C: at high $\bar{g}_\mathrm{Kir}$ and low $\bar{g}_\mathrm{KM}$, $\overline{V}(I)$ within the bistability window ranged from $-82.6$ to $\qty{-64.6}{\milli\volt}$ (pink), whereas at low $\bar{g}_\mathrm{Kir}$ and high $\bar{g}_\mathrm{KM}$, it shifted to $-64.6$ to $\qty{-56.4}{\milli\volt}$ with a smaller $\Delta I$ (purple). Increasing $\bar{g}_\mathrm{Kir}$ in the latter condition slightly lowered $\overline{V}(I)$ to $-67$ to $\qty{-57.6}{\milli\volt}$, despite a small reduction in $\Delta I$ of $\qty{0.08}{\micro\ampere\per\centi\meter^2}$ (yellow). A similar analysis at half the baseline $\bar{p}_{\mathrm{CaL}}$ (see \nameref{app:altgt}) showed that while Kir and KM channels promote distinct ranges of resting equilibria, KM channels can slightly increase $\Delta I$ when $\Delta I$ is too small to be considered robust ($\Delta I_0 = 0.34 < \qty{1.6}{\micro\ampere\per\centi\meter^2}$; largest increase: $\qty{0.14}{\micro\ampere\per\centi\meter^2}$ at $\bar{g}_\mathrm{Kir}=0$). Altogether, the effects observed in isolation persisted when both channels were combined: Kir channels promote robust bistability by enlarging $\Delta I$, whereas KM channels generally slightly reduce it.

\subsection*{The effects of CaL and Kir channels on bistability remain in the presence of additional conductances}

To test whether these effects generalize beyond the minimal model, we repeated the analysis on a published two-compartment model of deep projection neurons that already includes CaL channels in the dendrite and Kir channels in the soma, alongside calcium-dependent potassium (KCa) channels in both compartments, calcium-dependent nonspecific cation (CAN) channels in the dendrite, and a faster L-type calcium current ($\mathrm{CaL,f}$) in the soma~\cite{le_franc_multiple_2010}. We varied only $\bar{p}_\mathrm{CaL}$ and $\bar{g}_\mathrm{Kir}$ and kept the other conductances at their published values.

\begin{figure}[t!]
    \begin{center}
        \includegraphics[width=\textwidth]{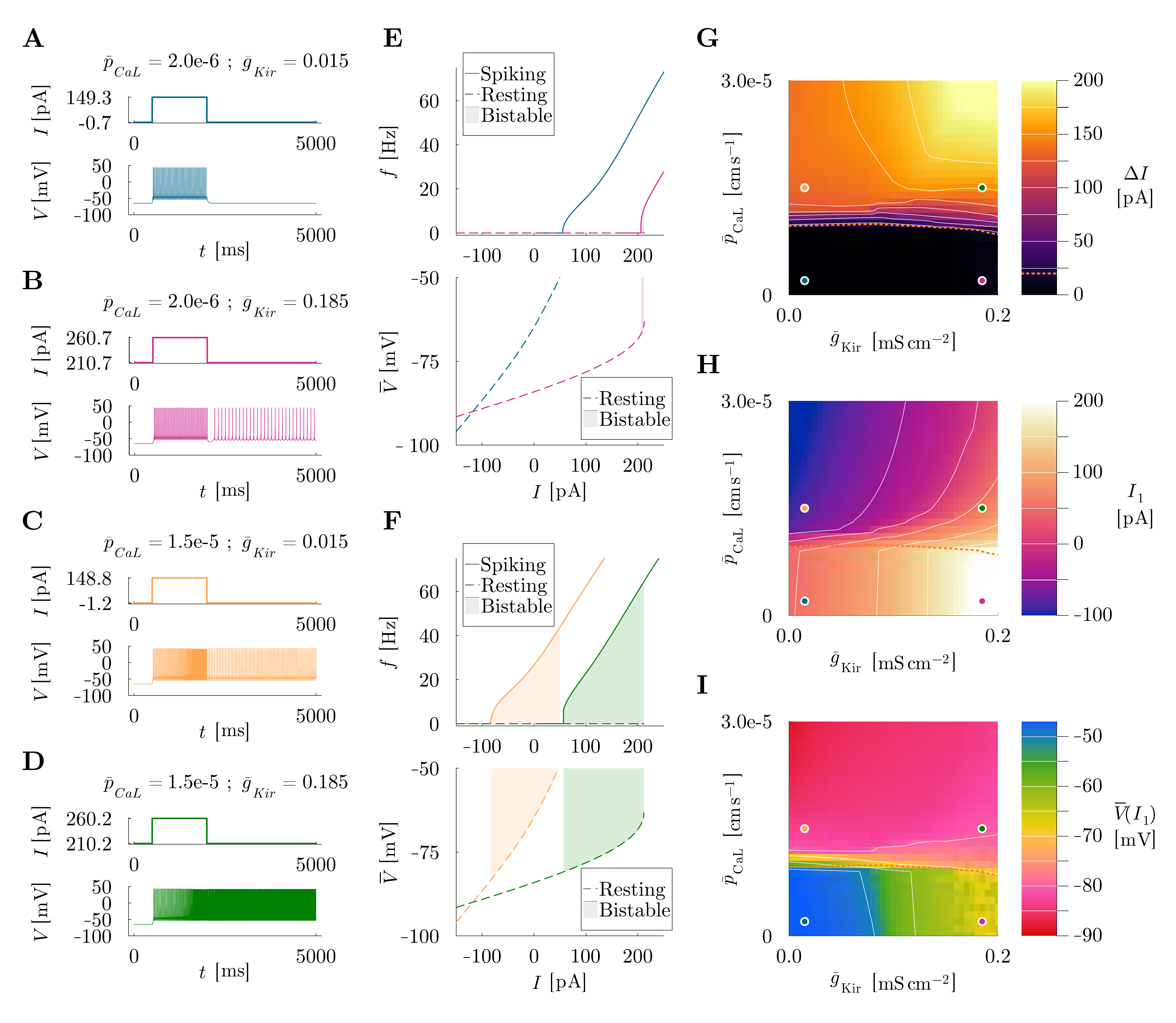} 
        \caption{{%
        In a two-compartment model of deep projection neurons including additional ion channels, CaL and Kir channels retain their effects on bistability.
        A:~Response of the model with low $\bar{p}_\mathrm{CaL}=\qty{2e-6}{\centi\meter\per\second}$ and low $\bar{g}_\mathrm{Kir}=\qty{0.015}{\milli\siemens\per\centi\meter^2}$, initially at $\qty{-65}{\milli\volt}$, to a $\qty{150}{\pico\ampere}$ current pulse.
        B:~Same as A with $\bar{g}_\mathrm{Kir}$ raised to $\qty{0.185}{\milli\siemens\per\centi\meter^2}$ and a $\qty{50}{\pico\ampere}$ pulse.
        C:~Same as A with $\bar{p}_\mathrm{CaL}$ raised to $\qty{1.5e-5}{\centi\meter\per\second}$.
        D:~Same as C with $\bar{g}_\mathrm{Kir}$ also raised to $\qty{0.185}{\milli\siemens\per\centi\meter^2}$ and a $\qty{50}{\pico\ampere}$ pulse.
        E:~fI curves (top) and corresponding resting equilibria (bottom) for the parameter sets used in A and C, isolating the effect of raising $\bar{p}_\mathrm{CaL}$.
        F:~Same as E for the parameter sets used in B and D.
        G--I:~Heatmaps of the bistability window size $\Delta I$ (G), its lower limit $I_1$ (H), and the corresponding resting equilibrium $\overline{V}(I_1)$ (I), as a function of $\bar{p}_\mathrm{CaL}$ and $\bar{g}_\mathrm{Kir}$. Each heatmap contains $41 \times 31$ data points. White lines are contour lines. 
        The red dashed line marks $\Delta I = \qty{20}{\pico\ampere}$, the robustness threshold estimated from published experimental data of afterdischarges. All parameter combinations shown in G--I satisfy $\overline{V}(I_1) \geq \qty{-90}{\milli\volt}$, so robust and physiological bistability coincides with the region above the red dashed line.
        {\label{fig:comp}} }}
    \end{center}
\end{figure}

We first examined the response of four illustrative pairs $(\bar{p}_\mathrm{CaL}, \bar{g}_\mathrm{Kir})$ to pulses of current of either $50$ or $\qty{150}{\pico\ampere}$, each with a baseline current producing a resting equilibrium of $\qty{-65}{\milli\volt}$. At low values of both conductances ($\bar{p}_{\mathrm{CaL}} = \qty{2e-6}{\centi\meter\per\second}$ and $\bar{g}_{\mathrm{Kir}}= \qty{0.015}{\milli\siemens\per\centi\meter^2}$), no bistability was observed (Fig.~\ref{fig:comp}A). Raising $\bar{g}_\mathrm{Kir}$ alone to $\qty{0.185}{\milli\siemens\per\centi\meter^2}$ (Fig.~\ref{fig:comp}B), $\bar{p}_\mathrm{CaL}$ alone to $\qty{1.5e-5}{\centi\meter\per\second}$ (Fig.~\ref{fig:comp}C), or both together (Fig.~\ref{fig:comp}D) all produced bistability.
The fI curves resulting from the first and the second pairs of conductances (used in Fig.~\ref{fig:comp}A and B, respectively), each with $\bar{p}_{\mathrm{CaL}} = \qty{2e-6}{\centi\meter\per\second}$, did not display a significant bistability window, as only the second pair of conductance exhibited a small $\Delta I$ of $\qty{6.3}{\pico\ampere}$ (Fig.~\ref{fig:comp}E, top). Increasing $\bar{g}_{\mathrm{Kir}}$ hyperpolarized $\overline{V}(I_2)$ from $-47$ to $\qty{-63}{\milli\volt}$ and increased $I_1$ (Fig.~\ref{fig:comp}E, bottom). With $\bar{p}_{\mathrm{CaL}}$ raised to $\qty{1.5e-5}{\centi\meter\per\second}$, $\Delta I$ increased to $\qty{133}{\pico\ampere}$ and $\qty{151.1}{\pico\ampere}$ for $\bar{g}_{\mathrm{Kir}} = \qty{0.015}{}$ and $\qty{0.185}{\milli\siemens\per\centi\meter^2}$, respectively, while increasing $\bar{g}_{\mathrm{Kir}}$ again hyperpolarized $\overline{V}(I_2)$ from $-48$ to $\qty{-63}{\milli\volt}$ and increased $I_1$ (Fig.~\ref{fig:comp}F).
The difference in pulse amplitude across Fig.~\ref{fig:comp}A--D ($50$ vs $\qty{150}{\pico\ampere}$) reflects the fact that, at high $\bar{g}_\mathrm{Kir}$, a current of only $\qty{10}{\pico\ampere}$ was sufficient to cross $I_1$ from a resting equilibrium of $\qty{-65}{\milli\volt}$.

A systematic sweep over the two conductances reproduced the minimal model pattern (Fig.~\ref{fig:comp}G): $\Delta I$ grew with $\bar{p}_\mathrm{CaL}$, and, for $\bar{p}_\mathrm{CaL}$ between $\qty{1.5e-5}{}$ and $\qty{3e-5}{\centi\meter\per\second}$, also with $\bar{g}_\mathrm{Kir}$. 
The increase in $\Delta I$ with $\bar{p}_{\mathrm{CaL}}$ appeared to saturate for $\bar{g}_{\mathrm{Kir}} \leq \qty{0.1}{\milli\siemens\per\centi\meter^2}$. Doubling $\bar{p}_{\mathrm{CaL}}$ from $\qty{1.5e-5}{}$ to $\qty{3e-5}{\centi\meter\per\second}$ increased $\Delta I$ by only $\qty{4}{\pico\ampere}$ at $\bar{g}_{\mathrm{Kir}} = 0$ and $\qty{21}{\pico\ampere}$ at $\bar{g}_{\mathrm{Kir}} = \qty{0.1}{\milli\siemens\per\centi\meter^2}$. Above this value, $\Delta I$ became more sensitive to $\bar{p}_{\mathrm{CaL}}$, reaching an increase of $\qty{55}{\pico\ampere}$ at $\bar{g}_{\mathrm{Kir}} = \qty{0.2}{\milli\siemens\per\centi\meter^2}$. 
At lower $\bar{p}_{\mathrm{CaL}}$ (between $\qty{1e-5}{}$ and $\qty{1.5e-5}{\centi\meter\per\second}$), $\Delta I$ still increased overall with $\bar{g}_{\mathrm{Kir}}$, but exhibited a transient drop of up to $\qty{25}{\percent}$ relative to its value at $\bar{g}_{\mathrm{Kir}}=0$ before rising again.

$I_1$ and $\overline{V}(I_1)$ followed the same patterns as in the minimal model (Fig.~\ref{fig:comp}H,I). Raising $\bar{p}_\mathrm{CaL}$ lowered $I_1$ and hyperpolarized $\overline{V}(I_1)$, while raising $\bar{g}_\mathrm{Kir}$ raised $I_1$. The effect of $\bar{g}_\mathrm{Kir}$ on $\overline{V}(I_1)$ depended on whether bistability was already established. With a well-developed bistability window ($\bar{p}_\mathrm{CaL}$ between $\qty{1.5e-5}{}$ and $\qty{3e-5}{\centi\meter\per\second}$), increasing $\bar{g}_\mathrm{Kir}$ depolarized $\overline{V}(I_1)$ by up to $\qty{5}{\milli\volt}$. Where bistability was weak or absent ($\bar{p}_\mathrm{CaL} < \qty{1e-5}{\centi\meter\per\second}$, $\Delta I < \qty{30}{\pico\ampere}$), it hyperpolarized $\overline{V}(I_1)$ by up to $\qty{25}{\milli\volt}$. At intermediate $\bar{p}_\mathrm{CaL}$ values, the effect transitioned from hyperpolarization to depolarization as bistability emerged.

These results confirm that the minimal model findings generalize to a more realistic setting: CaL channels create bistability, Kir channels enlarge the window, and both effects persist in the presence of KCa, CAN, and $\mathrm{CaL,f}$ currents. Accounting for the soma surface area ($\qty{1e-4}{\centi\meter^2}$), a $\qty{200}{\pico\ampere}$ window in the two-compartment model corresponds to $\qty{2}{\micro\ampere\per\centi\meter^2}$ in the minimal model. 
Although KM channels were not originally included in the two-compartment model, replacing Kir channels with KM channels in a full parameter sweep reduced $\Delta I$, consistent with the minimal model results, and in some cases eliminated bistability altogether (see \nameref{app:KM_in_comp}). Having established that the effects generalize, we next ask whether the two pairs also differ in their robustness to noise and intrinsic variability.


\subsection*{Bistability is more robust to noisy inputs and intrinsic variability when CaL and Kir channels are combined}

We next asked whether the bistabilities produced by CaL+KM and CaL+Kir pairs differ in robustness to noisy inputs and to intrinsic variability. To compare the two pairs on equal footing, we selected a CaL+KM set and a CaL+Kir set with matching bistability window sizes; because KM channels reduce $\Delta I$ while Kir channels enlarge it, this required a higher $\bar{p}_\mathrm{CaL}$ for the CaL+KM pair, while $\bar{g}_\mathrm{KM}$ and $\bar{g}_\mathrm{Kir}$ were set to the same value.

\begin{figure}
    \begin{center}
        \includegraphics[width=\textwidth]{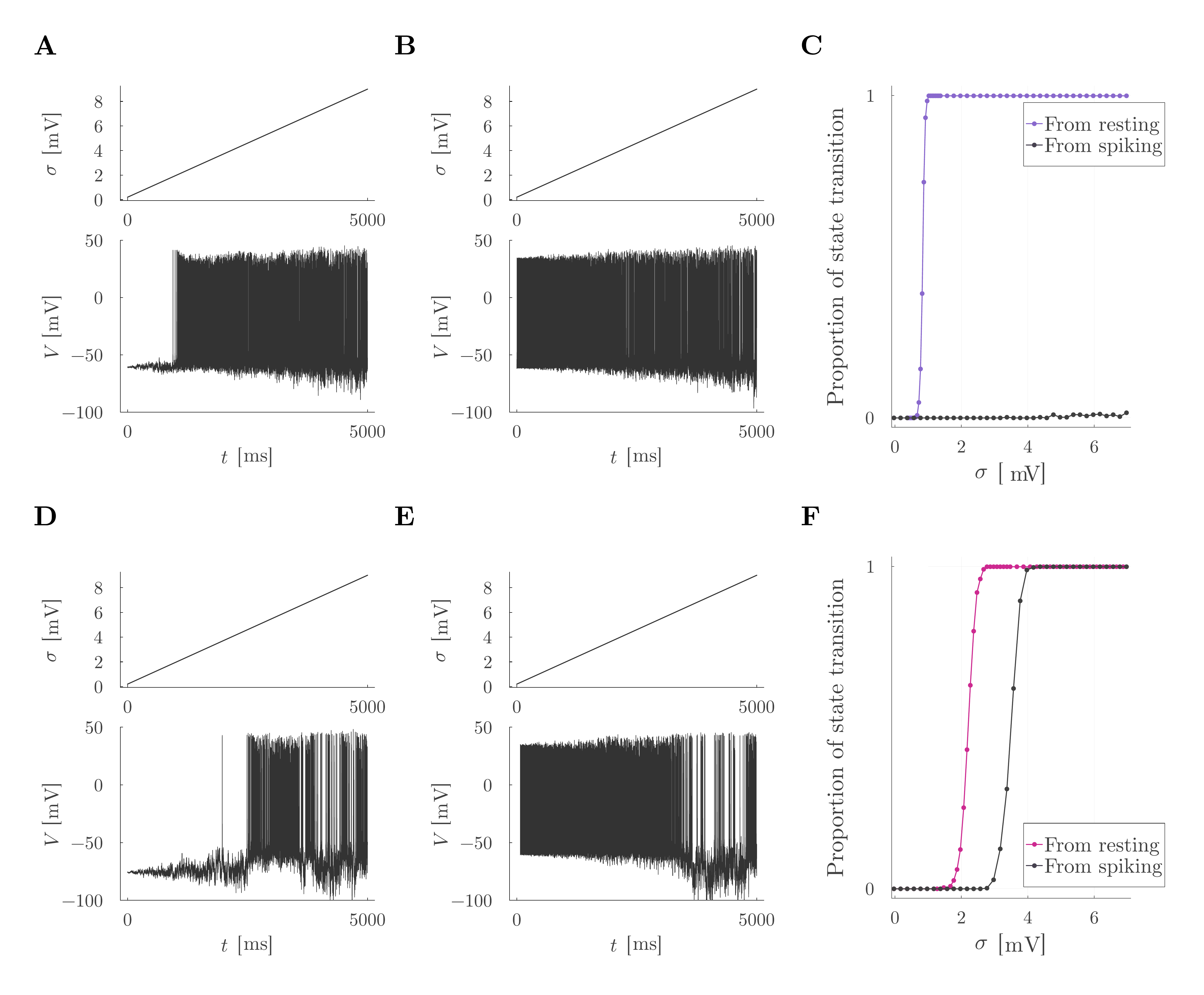} 
        \caption{{%
        Resting and spiking show different robustness to noise depending on whether KM or Kir channels are combined with CaL channels.
        A:~Membrane potential of the CaL+KM model in response to the central current of the bistability window, with superimposed white noise of increasing standard deviation $\sigma$, starting from the resting state.
        B:~Same as A, starting from the spiking state.
        C:~Proportion of 500 trials that transitioned away from the initial state (resting$\to$spiking or spiking$\to$resting) as a function of $\sigma$ held constant throughout each trial, for the CaL+KM model.
        D--F:~Same as A--C for the CaL+Kir model. The two models were matched for bistability window size: $\bar{g}_\mathrm{KM}$ (A--C) and $\bar{g}_\mathrm{Kir}$ (D--F) were set to the same value, while $\bar{p}_\mathrm{CaL}$ was higher in the CaL+KM model to compensate for the narrowing effect of KM channels.
        {\label{fig:4}} }}
    \end{center}
\end{figure}

    With CaL+KM channels, the resting state was lost under small perturbations while spiking was maintained up to the largest noise levels tested. Applying the central current of the bistability window with superimposed white noise of increasing standard deviation $\sigma$, the initially resting neuron transitioned to spiking within the first few tens of seconds (Fig.~\ref{fig:4}A), whereas the initially spiking neuron kept spiking (Fig.~\ref{fig:4}B). Across 500 trials at each fixed $\sigma$, the proportion of resting-to-spiking transitions grew rapidly with $\sigma$, while spiking-to-resting transitions never occurred in our range (Fig.~\ref{fig:4}C).

    With CaL+Kir channels, both states could be lost under sufficient noise, and the two states were of comparable robustness. The initially resting neuron transitioned to spiking at moderate $\sigma$ (Fig.~\ref{fig:4}D), and the initially spiking neuron transitioned to resting at slightly higher $\sigma$, with each resting episode lasting a few hundred milliseconds before spiking resumed (Fig.~\ref{fig:4}E). Spiking was more robust than resting in $\bar{g}\mathrm{Kir}$-bistability, tolerating higher noise levels even when kept constant over time (Fig.~\ref{fig:4}F), in contrast to $\bar{g}\mathrm{KM}$-bistability where resting was highly fragile while spiking remained undisrupted (Fig.~\ref{fig:4}C).

\begin{figure}
    \begin{center}
        \includegraphics[width=\textwidth]{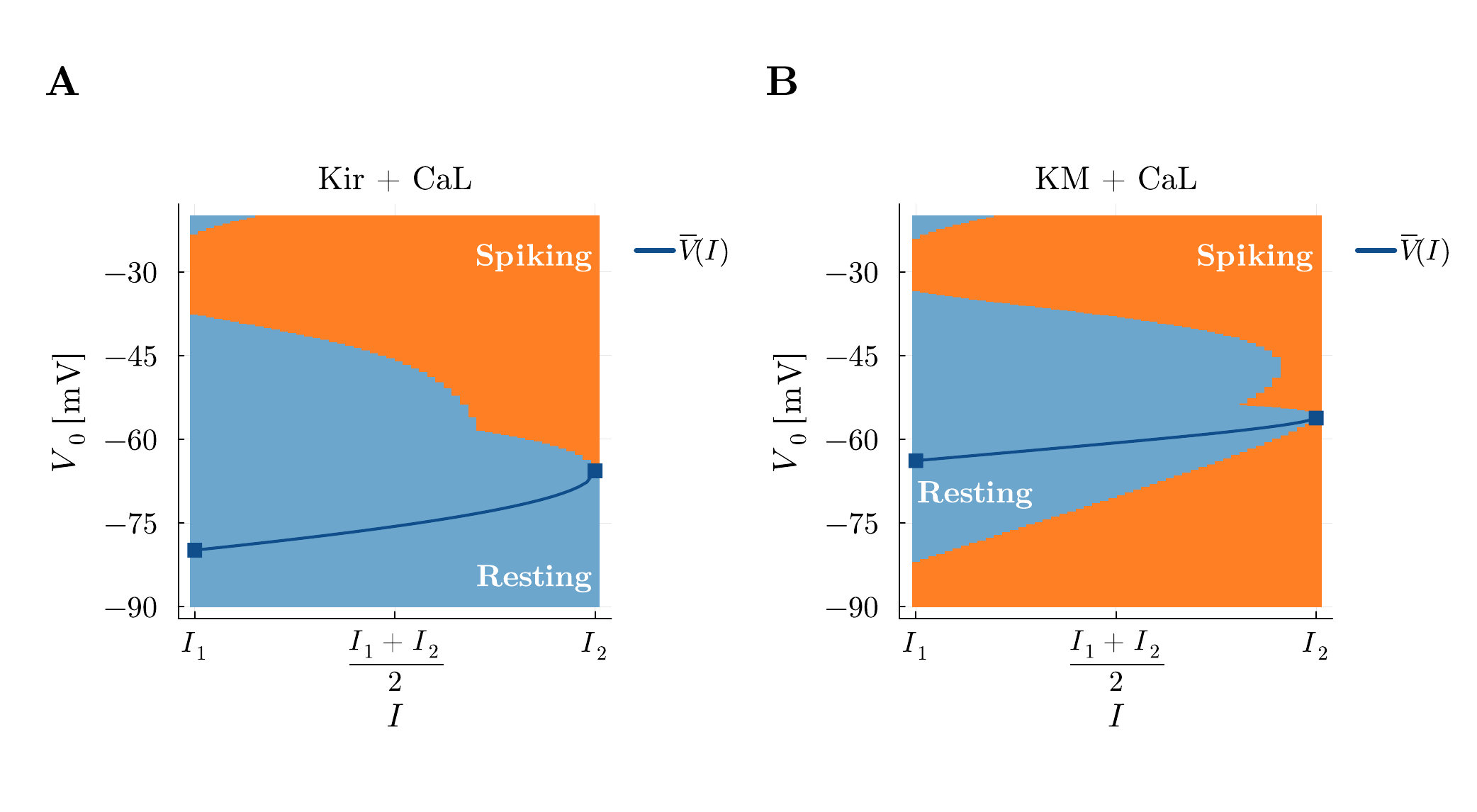} 
        \caption{{%
        The reachability of resting and spiking across the bistability window depends on which potassium channel is combined with CaL.
        A:~Final state reached by the CaL+Kir model as a function of the applied current (within $[I_1, I_2]$) and the initial membrane potential $V_0 \in [-90,-20]~\mathrm{mV}$. Blue indicates convergence to resting, orange to spiking. The dark blue line traces $\overline{V}(I)$ within the bistability window.
        B:~Same as A for the CaL+KM model.
        {\label{fig:6B}} }}
    \end{center}
\end{figure}

    This asymmetry in robustness raises the question of whether both states are equally reachable, particularly in $\bar{g}_\mathrm{KM}$-bistability where resting may be almost unreachable. To address this, we simulated the neuron response to a constant current within the bistability window (i.e., between $I_1$ and $I_2$) in noise-free conditions, for both $\bar{g}_\mathrm{Kir}$ and $\bar{g}_\mathrm{KM}$-bistability, starting from initial conditions 
    \begin{center}
        $(V_0, m_{\mathrm{ion},\infty}(V_0), h_{\mathrm{ion},\infty}(V_0), \ldots, \left[\mathrm{Ca}^{2+}\right]_0)$ with $V_0 \in [-90, -20]~\mathrm{mV}$, 
    \end{center}
    and recorded the state to which the model converged (Fig.~\ref{fig:6B}). 
    With CaL+Kir channels (Fig.~\ref{fig:6B}A), resting was more reachable than with CaL+KM channels (Fig.~\ref{fig:6B}B), the latter showing lower reachability of resting throughout the bistability window, while spiking showed the opposite trend. These results were consistent with those in Fig.~\ref{fig:4}, obtained at the center of each bistability window. Near $I_2$, the contrast was particularly striking: in $\bar{g}_\mathrm{KM}$-bistability, only initial conditions close to $\overline{V}(I_2)$ converged to resting, making this state extremely difficult to reach and maintain.
    These results also showed that state reachability varied within each bistability window, suggesting that noise robustness should similarly depend on the applied current. Consistently, repeating the experiment of Fig.~\ref{fig:4} at $I = I_1$ instead of $(I_1+I_2)/2$ showed that resting became more robust than spiking in both bistability types, while resting in $\bar{g}_\mathrm{Kir}$-bistability remained more robust than in $\bar{g}_\mathrm{KM}$-bistability (see~\nameref{app:noise_I1}).

\begin{figure}[h!]
    \begin{center}
        \includegraphics[width=\textwidth]{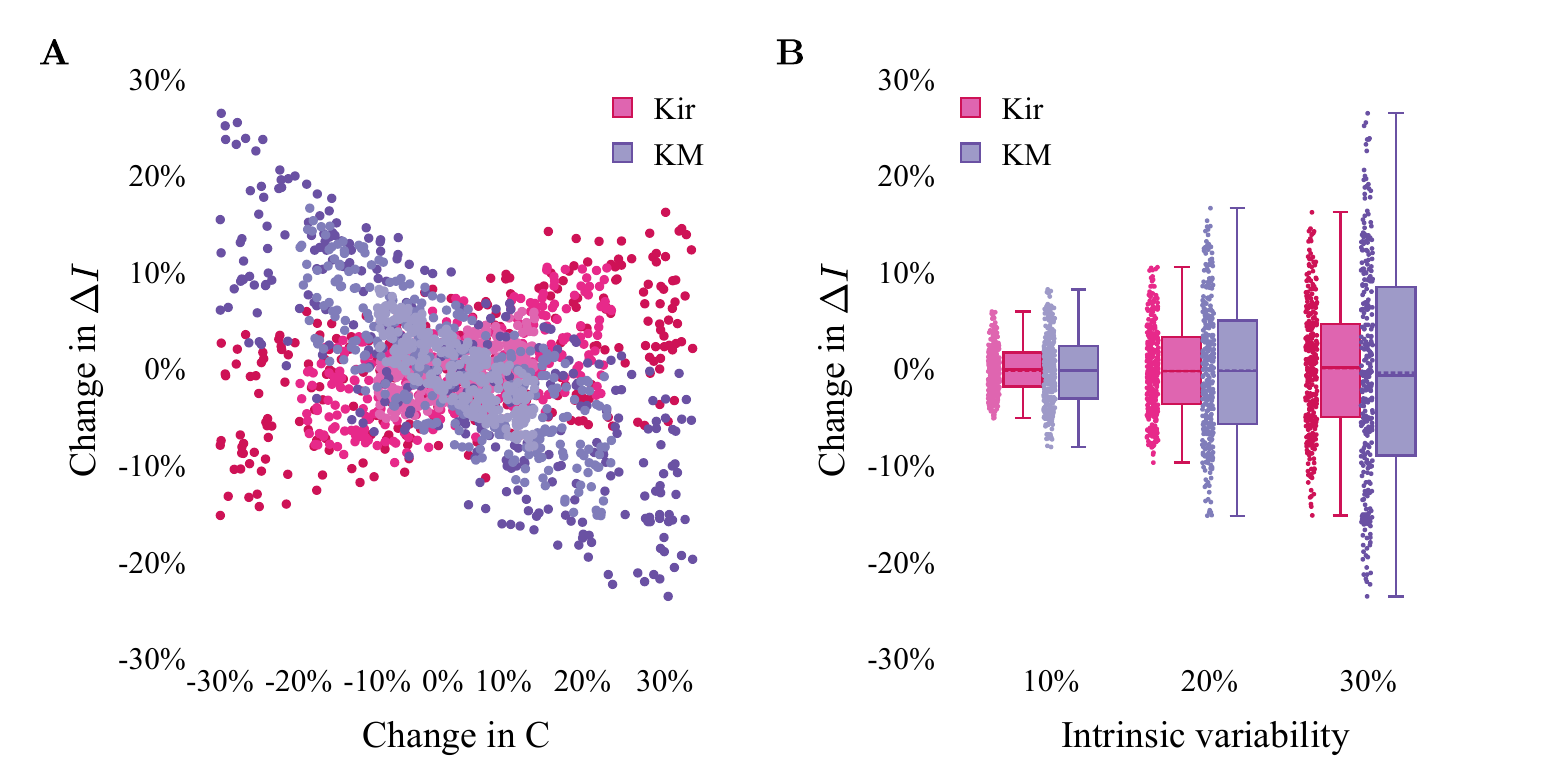}
        \caption{{%
        Intrinsic variability affects the bistability window size more strongly in CaL+KM than in CaL+Kir bistability.
        A:~Relative change in bistability window size $(\Delta I)$ as a function of the sampled membrane capacitance $C$, for the CaL+Kir (pink) and CaL+KM (purple) models. Shades of each color correspond to variability levels of $\pm 10\%$, $\pm 20\%$, and $\pm 30\%$ applied simultaneously to $C$, the sodium maximal conductance, and the leak conductance.
        B:~Boxplots of the relative changes in $\Delta I$ at each variability level, for the two models.
        {\label{fig:5}} }}
    \end{center}
\end{figure}

    We next investigated how both bistability types responded to intrinsic variability, estimating how $\Delta I$ evolved across heterogeneous neurons. The membrane capacitance ($C$), maximal sodium conductance ($\bar{g}_{\mathrm{Na}}$), and leak conductance ($g_{\mathrm{leak}}$) were each sampled uniformly within $\pm10\%$, $\pm20\%$, or $\pm30\%$ of their nominal values. The same conductance values were used for both bistability types ($\bar{g}_\mathrm{Kir} = \bar{g}_\mathrm{KM} = \qty{0.3}{\milli\siemens\per\centi\meter^2}$, $\bar{p}_\mathrm{CaL} = \qty{1.5e-5}{\centi\meter\per\second}$), yielding different $\Delta I$. The resulting relative changes in $\Delta I$ were computed and represented as a function of $C$ for each variability level (Fig.~\ref{fig:5}A). Changes in $C$ had opposite effects on the two bistability types, but with smaller magnitude for $\bar{g}_\mathrm{Kir}$-bistability. At every level of intrinsic variability tested, relative changes in $\bar{g}_\mathrm{KM}$-bistability were approximately 1.5 times larger than in $\bar{g}_\mathrm{Kir}$-bistability, which remained within $\pm20\%$ (Fig.~\ref{fig:5}B).

\subsection*{Ion channels increasing bistability have a steady-state current exhibiting a region of negative differential conductance}

    The opposing effects of KM and Kir channels on $\Delta I$ must arise from the shapes of their steady-state currents, since both are voltage-gated potassium channels operating on a similar timescale. We therefore compared the steady-state currents of CaL, KM, and Kir channels, focusing on their differential conductance around the spike threshold (${\approx}\qty{-65}{\milli\volt}$).

\begin{figure}
    \begin{center}
        \includegraphics[width=\textwidth]{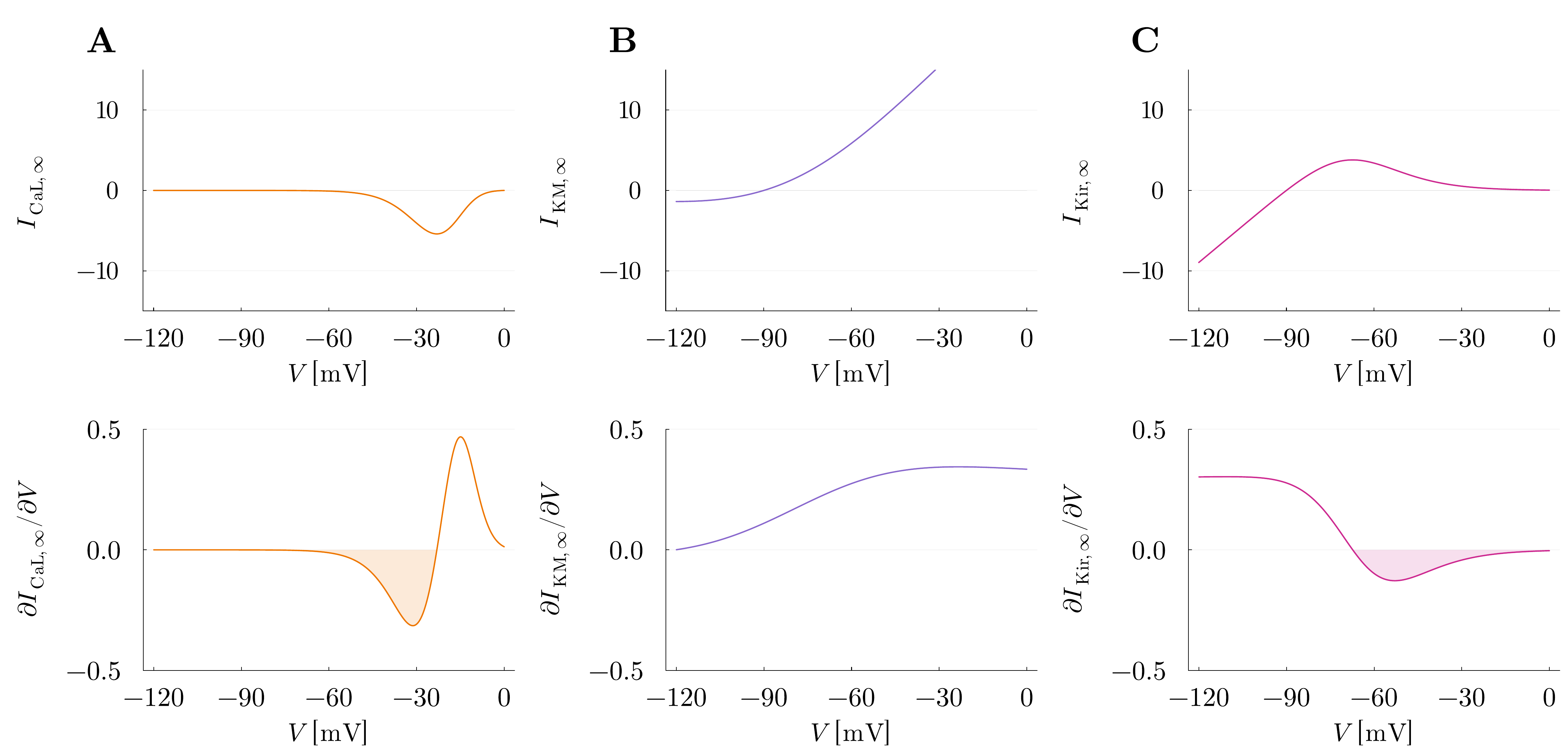}
        \caption{{%
        Steady-state currents of CaL, KM, and Kir channels and their differential conductance.
        A:~Steady-state L-type calcium current $I_{\mathrm{CaL},\infty}$ (top) and its differential conductance (bottom), at an intracellular calcium concentration of \SI{2}{\milli\molar}. The region of negative differential conductance around the spike threshold is shaded.
        B:~Same as A for the KM current $I_{\mathrm{KM},\infty}$, which has positive differential conductance throughout.
        C:~Same as A for the Kir current $I_{\mathrm{Kir},\infty}$, which has a region of negative differential conductance above $E_\mathrm{K}$ (shaded). The y-scale of the steady-state currents is in $\qty{}{\micro\ampere\per\centi\meter^2}$ (top) and that of the differential conductance in  $\qty{}{\micro\ampere\per\milli \volt\centi\meter^2}$. A,B,C used respectively $\overline{p}_{\mathrm{CaL}}=\qty{1.5e-5}{\centi\meter\per\second}$, $\overline{g}_{\mathrm{KM}}=\qty{0.3}{\milli\siemens\per\centi\meter^2}$, and $\overline{g}_{\mathrm{Kir}}=\qty{0.3}{\milli\siemens\per\centi\meter^2}$.
        {\label{fig:6}} }}
    \end{center}
\end{figure}

    CaL channels carry an inward current  (Fig.~\ref{fig:6}A, top) with a region of negative differential conductance near the spike threshold (Fig.~\ref{fig:6}A, shaded area). Previous work has linked this feature to the onset of bistability~\cite{franci_balance_2013,drion_dynamic_2015}: a small depolarization within this range increases the inward calcium current, which depolarizes the membrane further and triggers spiking at a lower applied current, thereby reducing $I_1$. Because CaL channels barely affect the values of resting equilibria (see Fig.\ref{fig:1}C and \nameref{app:fI_hh} prior to the addition of CaL channels), resting at a smaller $I_1$ must be balanced by a stronger leak current, that dominates the sum of ionic currents below $\qty{-60.1}{\milli\volt}$ (the reversal potential of the leak current). This accounts for the hyperpolarization of $\overline{V}(I_1)$ observed in Figs.~\ref{fig:1}C and~\ref{fig:3}C.

    Unlike CaL channels, KM channels carry a small inward and a strong outward current (Fig.~\ref{fig:6}B, top), both with strictly positive differential conductance (Fig.~\ref{fig:6}B, bottom). Empirically, increasing $\bar{g}_\mathrm{KM}$ shifts both $I_1$ and $I_2$ upward, but raises $I_1$ more than $I_2$, thereby reducing $\Delta I$ (Figs~\ref{fig:1}C and \ref{fig:2}B--C). 
    This asymmetric shift follows from $I_1$ and $I_2$ being governed by different dynamical states: spiking at $I_1$ and resting at $I_2$. When initially spiking at $I_1$, the membrane potential is overall higher than the resting equilibrium at $I_2$. Since KM channels have positive differential conductance, they produce a stronger outward current in the spiking scenario at $I_1$ than in the resting scenario at $I_2$, requiring a larger compensating current increase at $I_1$. Additionally, increasing $\bar{g}_\mathrm{KM}$ elevates $\overline{V}(I)$ (Fig.~\ref{fig:2}C), as KM channels provide a non-negligible current at these potentials and, together with the leak current, dominate the sum of ionic currents at rest, causing $\overline{V}(I_1)$ and $\overline{V}(I_2)$ to rise with $I_1$ and $I_2$  (Fig.\ref{fig:3}C). For a fixed applied current (e.g., $I = \qty{-0.1}{\micro\ampere\per\centi\meter^2}$), KM channels hyperpolarize the resting equilibrium toward $E_K$ (from $\overline{V} = \qty{-60.9}{\milli\volt}$ in Fig.~\ref{fig:1}C to below $\qty{-75}{\milli\volt}$ in Fig.~\ref{fig:2}C).

    In contrast to KM channels, Kir channels carry a strong inward current below $E_\mathrm{K}$ and a weaker outward current (Fig.~\ref{fig:6}C, top), and share with CaL channels a region of negative differential conductance around the spike threshold (Fig.~\ref{fig:6}C, bottom). By analogy with the CaL current, this negative-slope region promotes the resting-to-spiking transition: a small depolarization decreases the outward Kir current, which depolarizes the membrane further. Like KM channels, increasing $\bar{g}_\mathrm{Kir}$ shifts both $I_1$ and $I_2$ upward, but the shift is larger for $I_2$ than for $I_1$, in contrast to KM channels, thereby increasing $\Delta I$ (Figs~\ref{fig:1}C and \ref{fig:2}E--F). By the same reasoning as for KM channels, Kir channels produce a lower outward current in the spiking scenario at $I_1$ than in the resting scenario at $I_2$, requiring a larger compensating current increase at $I_2$ because of the negative differential conductance of the outward Kir current. 
    Increasing $\bar{g}_\mathrm{Kir}$ generally elevates $\overline{V}(I)$ (Fig.~\ref{fig:2}F), as Kir channels dominate the sum of ionic currents at rest, though their effect is non-linear. Below $\qty{-67}{\milli\volt}$, Kir channels display positive differential conductance, which steepens the sum of ionic currents at rest. In this region, the rise in $I_1$ induced by increasing $\bar{g}_\mathrm{Kir}$ therefore depolarizes $\overline{V}(I_1)$ (Figs~\ref{fig:2}F and \ref{fig:3}F, upper half). Above $\qty{-67}{\milli\volt}$, the Kir current increasingly dominates the sum of ionic currents as the potential approaches $\qty{-60.1}{\milli\volt}$, since the leak current diminishes near its reversal potential. In this region, however, the Kir current displays negative differential conductance, creating a negative-slope region in the sum of ionic currents near $\qty{-60.1}{\milli\volt}$. To understand the contribution of Kir channels to $\overline{V}(I_2)$, consider a neuron with the same conductances as in Fig.~\ref{fig:2}F, initially resting at $I_1$, with the applied current progressively increased. As the membrane potential rises, the positive differential conductance of the Kir current initially allows the sum of ionic currents to compensate for the depolarization. This compensation reaches its limit at $\delta (\Sigma I_\mathrm{ion}) /\delta V = 0$, just before the negative-slope region of the sum of ionic currents, determining $\overline{V}(I_2)$. As $\bar{g}_\mathrm{Kir}$ increases, the negative-slope region grows, shifting this limit to more hyperpolarized potentials and thus progressively hyperpolarizing $\overline{V}(I_2)$, as observed when $I_2=I_1$ for $\Delta I=0$ and $\bar{p}_{\mathrm{CaL}}\sim 0$ in Fig.~\ref{fig:3}F.

    Both KM and Kir channels produce an asymmetric shift in $I_1$ and $I_2$,  which respectively narrows or enlarges $\Delta I$. This asymmetry arises because $I_1$ and $I_2$ are set by the appearance and disappearance of two different behaviors, governed by distinct dynamical mechanisms that rely on different types of bifurcations (examined later, Fig.~\ref{fig:8}). These mechanisms are potentially shaped by the sign of the differential conductance around the spike threshold, suggesting that its sign predicts whether a potassium channel enlarges or narrows $\Delta I$. We test this hypothesis in the next subsection by selectively blocking the inward and outward components of each current.

\subsection*{The inward currents redress resting equilibria and maintain bistability, but the outward potassium currents either increase or decrease bistability}

    If the sign of the differential conductance around the spike threshold dictates the effect on $\Delta I$, then only the outward component of each potassium current, which carries that differential conductance, should reshape the bistability window, while the inward component, being zero above $E_\mathrm{K}$, should not interfere with spiking. We tested this by blocking either the inward or the outward component of each current in turn, measuring the resulting $\Delta I$ and the resting equilibria $\overline{V}(I)$ within the bistability window, the latter being determined by applying hyperpolarizing current steps between $I_1$ and $I_2$.

\begin{figure}
    \begin{center}
        \includegraphics[width=\textwidth]{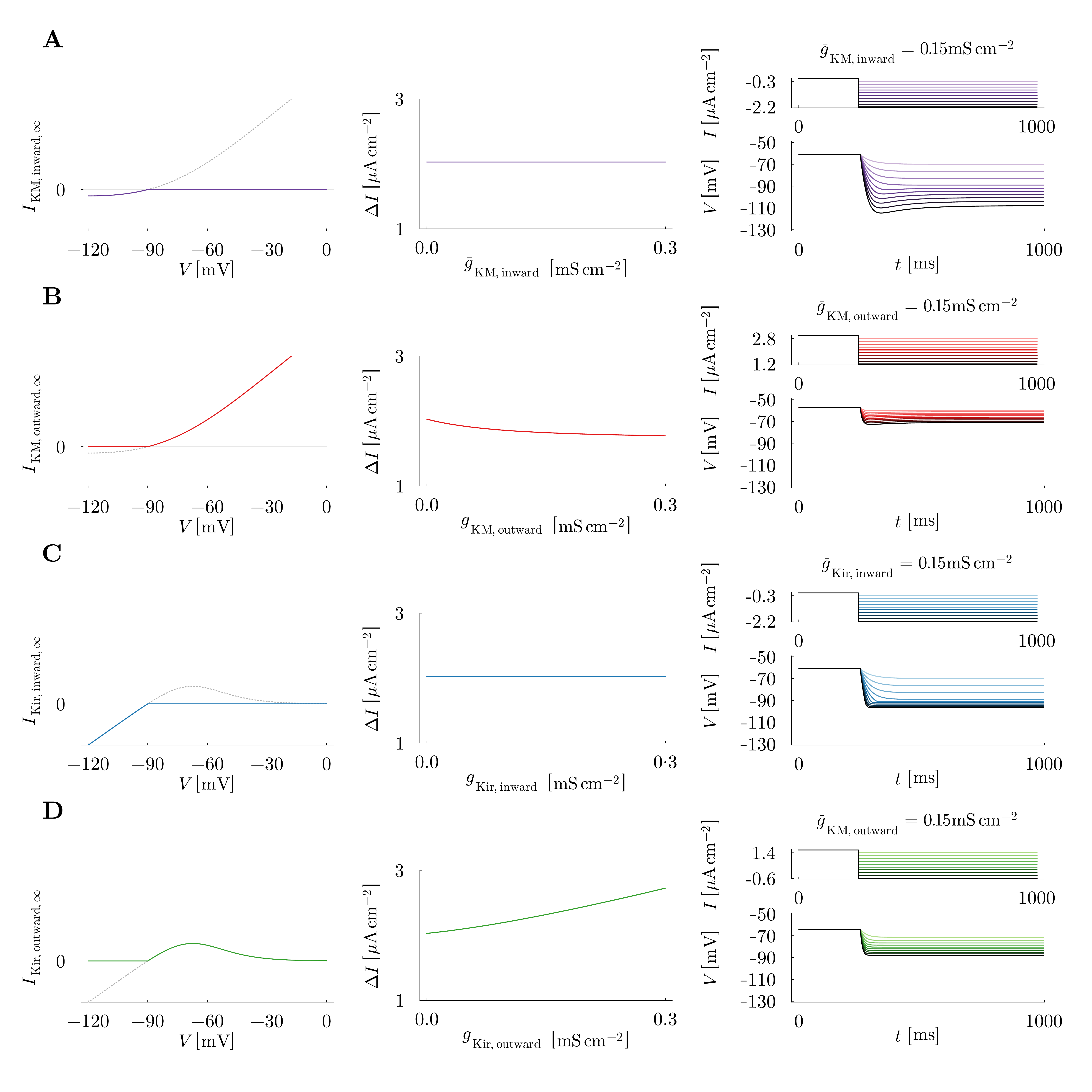}
        \caption{{%
        Inward components of both currents redress resting equilibria below $\qty{-90}{\milli\volt}$, while outward components reshape the bistability window in opposite directions.
        A:~Steady-state KM current with the outward component blocked (left), bistability window size $\Delta I$ as a function of the inward conductance $\bar{g}_\mathrm{KM,inward}$ (center), and hyperpolarizing steps of current across the window at a fixed $\bar{g}_\mathrm{KM,inward}$ (right).
        B:~Same as A with the inward component of the KM current blocked instead.
        C:~Same as A for the Kir current, with the outward component blocked.
        D:~Same as A for the Kir current, with the inward component blocked. In the right-hand panels, the maximum and minimum applied currents equal $I_1$ and $I_2$ for each blocked condition, at a conductance of $\qty{0.15}{\milli\siemens\per\centi\meter^2}$.
        {\label{fig:7}}  }}
    \end{center}
\end{figure}

    To isolate the inward KM current, the outward component was blocked above $E_\mathrm{K}$ (Fig.~\ref{fig:7}A, left). With only its inward component active, the KM current did not alter $\Delta I$ (Fig.~\ref{fig:7}A, center). Hyperpolarizing steps within the bistability window showed that it only slightly raised the resting equilibria that lay below $E_\mathrm{K}$ (Fig.~\ref{fig:7}A, right). Conversely, isolating the outward KM current by blocking the inward component (Fig.~\ref{fig:7}B, left) showed that increasing $\bar{g}_\mathrm{KM, outward}$ narrowed $\Delta I$ while efficiently adjusting resting equilibria (Fig.~\ref{fig:7}B, center--right).

    Blocking the outward Kir current isolated the inward component (zero above $E_\mathrm{K}$, Fig.~\ref{fig:7}C, left), which left $\Delta I$ unchanged as $\bar{g}_\mathrm{Kir, inward}$ increased (Fig.~\ref{fig:7}C, center), but efficiently hyperpolarized resting equilibria up to $E_\mathrm{K}$ (Fig.~\ref{fig:7}C, right). This correction was more efficient than for the inward KM current, due to the larger amplitude of the inward Kir current. Conversely, blocking the inward Kir current isolated the outward component (non-zero above $E_\mathrm{K}$, Fig.~\ref{fig:7}D, left), which enlarged the bistability window as $\bar{g}_\mathrm{Kir, outward}$ increased, while only adjusting resting equilibria above $E_\mathrm{K}$ (Fig.~\ref{fig:7}D, center--right).

    These four experiments together show a clear partition of roles. The inward components of both KM and Kir currents act only on resting equilibria below $E_\mathrm{K}$ and leave $\Delta I$ unchanged, since the spike threshold lies above $E_\mathrm{K}$ and the inward currents vanish there. The outward components redress the remaining resting equilibria and, in addition, act on $\Delta I$: the outward KM current narrows it while the outward Kir current enlarges it. This difference is consistent with the sign of their differential conductance around the spike threshold, which is positive for KM and negative for Kir, confirming the hypothesis of the previous subsection.

\subsection*{CaL and Kir channels facilitate the generation of plateau potentials, whereas KM channels induce a distinct excitability switch}

    The qualitative changes in $\overline{V}(I_1)$ and $\Delta I$ observed as $\bar{p}_\mathrm{CaL}$, $\bar{g}_\mathrm{Kir}$, or $\bar{g}_\mathrm{KM}$ increases (Fig.~\ref{fig:3}) suggest that each channel type triggers a distinct excitability switch, that is, a qualitative change in how the neuron approaches or leaves the spiking regime. To identify these switches, we used the minimal conductance-based model containing only $I_\mathrm{Na}$, $I_\mathrm{KDR}$, and $I_\mathrm{leak}$ as a reference, and compared it to three variants in which CaL, Kir, or KM channels were added individually. For each condition we recorded the response to a hyperpolarizing current step crossing $I_1$ (Fig.~\ref{fig:8}A--D) and computed the bifurcation diagram of the membrane potential as a function of the applied current (Fig.~\ref{fig:8}E--H).

    In the reference condition ($\bar{p}_\mathrm{CaL}=\bar{g}_\mathrm{Kir} = \bar{g}_\mathrm{KM}=0$), reducing the applied current below $I_1$ causes the neuron to settle directly to a resting equilibrium within the range of potentials spanned by spiking (Fig.~\ref{fig:8}A). 
    The bifurcation diagram confirmed this, showing that resting equilibria near $I_1$ exceeded the minimum of the spiking limit cycle (Fig.~\ref{fig:8}E). The overlap between $I_1$ and $I_2$ reflects the absence of bistability, as spiking appeared precisely when resting disappeared. As the applied current increased, the resting state disappeared at $I_2$ through a saddle-node (SN) bifurcation, where the stable equilibrium collided with a saddle point. Since the stable equilibrium at $I_2$ lay within the spiking voltage range and spiking frequency near $I_1$ was of the order of a few hertz (Fig.~\ref{fig:8}A), the SN bifurcation occurred on an invariant circle, giving rise to a saddle-node on invariant circle (SNIC) bifurcation. By nature, the SNIC bifurcation is associated with low-frequency spiking at onset \cite{franci_organizing_2012}.

\begin{figure}
    \begin{center}
        \includegraphics[width=\textwidth]{figures/fig-11.png} %
        \caption{{%
        Excitability switches created by CaL, Kir, and KM channels, identified through bifurcation analysis.
        A--D:~Response of the model to a hyperpolarizing current step crossing $I_1$, for four conditions. A: reference model ($I_\mathrm{Na}$, $I_\mathrm{KDR}$, $I_\mathrm{leak}$). B: reference model $+$ CaL channels. C: reference model $+$ Kir channels. D: reference model $+$ KM channels. In each panel the lower and upper current values are $I_1 \pm \qty{10}{\nano\ampere\per\centi\meter^2}$ for the corresponding conductances.
        E--H:~Bifurcation diagrams for the four conditions. Bifurcation types are labelled directly on each panel: SNIC (saddle-node on invariant circle), SN (saddle-node), SH (saddle-homoclinic), and Hopf.
        {\label{fig:8}} 
        }}
    \end{center}
\end{figure}

     Adding CaL channels alone ($\bar{p}_\mathrm{CaL} = \qty{7.5e-6}{\centi\meter\per\second}$) rendered the neuron bistable, and produced a plateau upon hyperpolarization below $I_1$ (Fig.~\ref{fig:8}B). This plateau represented a clear separation between the range of potentials encompassed by spiking and the resting equilibria below $I_1$, that are typically below the range of potentials covered by spiking. Bistability arose because the addition of CaL transformed the SNIC bifurcation into a separate SN bifurcation at $I_2$ and a saddle-homoclinic (SH) bifurcation at $I_1$ (Fig.~\ref{fig:8}F). The saddle-homoclinic bifurcation is defined by the collision between a saddle point and a homoclinic loop, which is defined by the high-dimensional model trajectory during spiking, at $I_1$, where spiking appears (for more details see  \cite{franci_organizing_2012}). However, the bifurcation diagram also revealed that the separation between resting equilibria and the spiking voltage range exists only near $I_1$, so the plateau observed upon hyperpolarization may not persist for currents closer to $I_2$ (Fig.~\ref{fig:8}F).

    Adding Kir channels alone ($\bar{g}_\mathrm{Kir} = \qty{0.3}{\milli\siemens\per\centi\meter^2}$) produced a smaller plateau (Fig.~\ref{fig:8}C) but generated the same qualitative bifurcation structure as CaL channels, transforming the SNIC bifurcation into a separate SN bifurcation and a SH bifurcation (Fig.~\ref{fig:8}G). This shared bifurcation structure likely underlies the enlarging effect of Kir channels on $\Delta I$. Notably, the voltage separation between resting equilibria and the spiking voltage range persisted throughout the entire bistability window, up to $I_2 = I_1$ (Fig.~\ref{fig:8}G).

     Adding KM channels alone ($\bar{g}_\mathrm{KM} = \qty{0.3}{\milli\siemens\per\centi\meter^2}$) produces a qualitatively different switch. Instead of a plateau, the hyperpolarizing step elicits damped subthreshold oscillations as the neuron approaches a higher resting equilibrium (Fig.~\ref{fig:8}D). The bifurcation diagram revealed that the resting equilibria largely overlapped the spiking voltage range (Fig.~\ref{fig:8}H). The resting state disappeared through a Hopf bifurcation rather than a SN bifurcation, as the stable equilibrium lost stability before colliding with the saddle point. The Hopf bifurcation confers resonator behavior, whereby the neuron preferentially responds to inputs at the resonant frequency of its subthreshold oscillations, in contrast to integrators associated with SN or SNIC bifurcations, which perform temporal integration of input pulses \cite{izhikevich_book}. Intuitively, this lack of voltage separation means that a neuron transitioning from spiking to rest remains within the voltage range visited during spiking, making the resting state far easier to escape than in the SN+SH case, where the transition requires crossing a voltage region not visited during spiking.

\begin{figure}[t!]
    \centering
    \includegraphics[width=\textwidth]{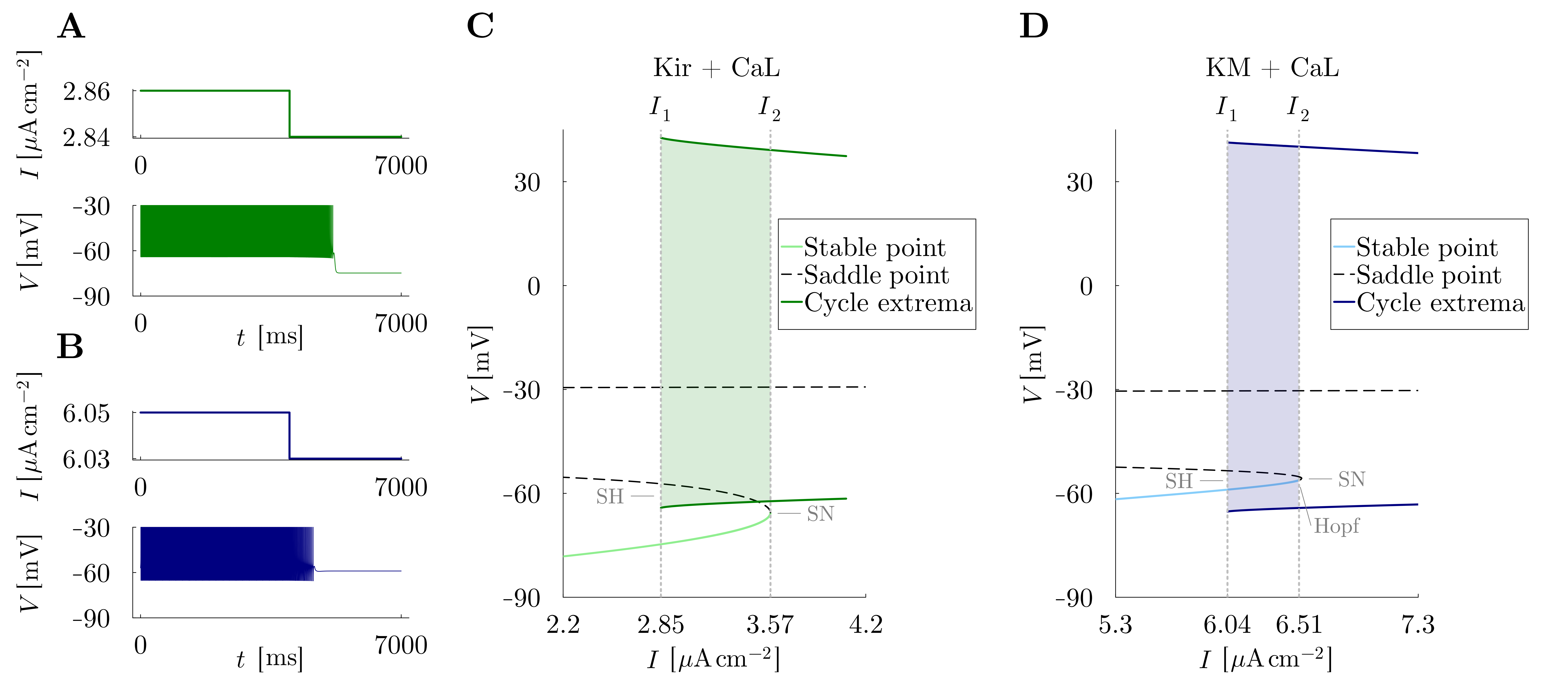} 
    \caption{%
        Combining CaL with Kir channels reinforces plateau potentials while combining CaL with KM channels suppresses them.
        A:~Response of the CaL+Kir model to a hyperpolarizing current step crossing $I_1$.
        B:~Same as A for the CaL+KM model. In A and B the lower and upper current values are $I_1 \pm \qty{10}{\nano\ampere\per\centi\meter^2}$ for the corresponding conductances. 
        The brief spiking following the step in B is an afterdischarge, not a plateau potential.
        C:~Bifurcation diagram of the CaL+Kir model as a function of the applied current.
        D:~Same as C for the CaL+KM model.
        Bifurcation types are annotated directly on the panel.
        }
    \label{fig:9-CaLs+Kir/KM}
\end{figure}

    When CaL and Kir channels are combined, the hyperpolarizing step produces a pronounced plateau (Fig.~\ref{fig:9-CaLs+Kir/KM}A), and the bifurcation diagram preserves the SN+SH structure observed for each channel alone (Fig.~\ref{fig:9-CaLs+Kir/KM}C). In contrast, combining CaL with KM channels suppresses the plateau (Fig.~\ref{fig:9-CaLs+Kir/KM}B), as resting equilibria are pulled into the spiking voltage range and disappear at $I_2$ through a Hopf bifurcation (Fig.~\ref{fig:9-CaLs+Kir/KM}D). The resonant behavior is less visible in Fig.~\ref{fig:9-CaLs+Kir/KM}B than in Fig.~\ref{fig:8}D because the hyperpolarizing step is applied at $I_1$, which lies further from the Hopf bifurcation than the current used in Fig.~\ref{fig:8}D. The two channel combinations thus differ not only in bistability window size and robustness, but also in the type of excitability they support: plateau-generating in CaL+Kir and tonic firing with resonant behavior in CaL+KM. 
    In CaL+Kir, the plateau directly reflects resting state robustness: the larger the voltage separation between resting and spiking, the harder it is for noise to drive the neuron out of rest and back into spiking. This separation is largest near $I_1$ and shrinks toward $I_2$, consistent with the current-dependent reachability and robustness observed in Fig.~\ref{fig:6B}A and Fig.~\ref{fig:4}C vs \nameref{app:fig:noise_I1}. In CaL+KM, the Hopf bifurcation eliminates this voltage separation and confers resonator behavior, both of which contribute to the lower robustness and reachability of the resting state compared to CaL+Kir, especially near the Hopf bifurcation (Figs~\ref{fig:4}, \ref{fig:6B}, and \nameref{app:fig:noise_I1}).
    The shared SH bifurcation at spiking onset in both cases also provides further insight into the asymmetric shift of $I_1$ and $I_2$ induced by Kir and KM channels. The spiking scenario discussed earlier depends on the membrane potential at which the saddle point lies, which is always higher than the resting equilibrium at $I_2$, supporting the conclusion that the outward Kir (resp. KM) current contribution at $I_1$ is lower (resp. higher) than at $I_2$, thereby increasing (resp. decreasing) $\Delta I$.

\section*{Discussion}

Cooperation between L-type calcium (CaL) channels and inward rectifier potassium (Kir) channels produces a bistability between resting and spiking states that is physiological, robust to noise and intrinsic variability, and capable of supporting plateau potentials and sustained afterdischarges. This combination provides a candidate intrinsic pathway by which deep dorsal horn projection neurons amplify nociceptive input during central sensitization. Our analysis also shows why this pair is special: Kir channels are unusual among voltage-gated potassium channels in that their outward steady-state current, like the CaL inward current, has a region of negative differential conductance around the spike threshold, the same shape feature that enlarges the bistability window.

Most voltage-gated potassium channels reduce the bistability window size in exchange for redressing the resting equilibria within the window, producing the trade-off that has made physiological bistability via CaL channels alone hard to explain~\cite{yuen_bistability_1995,franci_balance_2013,naudin_general_2023}. Kir channels avoid this trade-off through the negative slope of their outward current, which asymmetrically shifts the boundary behaviors of the bistability window to expand it, while anchoring the resting equilibria below the local negative-slope region that Kir channels introduce in the sum of ionic currents. The result is that the region of parameter space where bistability is both robust and physiological is larger for CaL+Kir than for CaL+KM, even though physiological bistability is achievable with either channel combination given sufficient CaL conductance. Selectively blocking either the inward or the outward Kir current reinforces this picture: the inward component exclusively redresses subthreshold resting equilibria, while the outward component alone is responsible for enlarging the bistability window. This clarifies a proposal by Amarillo and colleagues~\cite{amarillo_inward_2018} that Kir channels support intrinsic bistability, and generalizes it beyond thalamocortical neurons. The bifurcation analysis tells the same story. Both CaL and Kir channels produce a saddle-node $+$ saddle-homoclinic structure, with Kir channels shifting the saddle point outside the spiking limit cycle range and thereby creating a plateau that separates resting from spiking. KM channels, by contrast, drive the resting equilibrium through a Hopf bifurcation, preventing such a separation.

These functional differences translate into distinct modes of bistability with different implications for neuronal computation. In CaL+Kir bistability, resting and spiking are comparably robust to noise at the center of the bistability window, allowing a projection neuron to reliably hold either state and transition between them under stimulus control. CaL+KM bistability, by contrast, features robust spiking but fragile resting, so once the neuron begins spiking, it tends to remain in that state. This asymmetry is consistent with the Hopf-type loss of stability of the KM resting equilibrium and with our reachability analysis, which showed that the CaL+KM resting state near $I_2$ is accessible only from a narrow band of initial conditions, becoming increasingly fragile as the applied current approaches $I_2$. Near $I_1$, the resting state of CaL+KM becomes more robust than spiking, but both states show comparable robustness, so spiking remains generally dominant throughout the bistability window. For pain processing, where afterdischarges must eventually terminate to allow the system to return to baseline between stimuli, CaL+Kir is therefore the functionally more appropriate mechanism. This conclusion is further supported by our finding that CaL+Kir bistability is more reliable than CaL+KM bistability across a heterogeneous neuronal population exhibiting intrinsic bistability.

The occurrence of long-lasting afterdischarges depends critically on whether the baseline current between pulses falls within the bistability window. Modulatory inputs that shift the window boundaries can therefore gate afterdischarge occurrence: if the window shifts upward (e.g., due to increased $\bar{g}_\mathrm{Kir}$), a fixed baseline current may fall below $I_1$, preventing afterdischarges. Conversely, if the window shifts downward (e.g., due to decreased $\bar{g}_\mathrm{Kir}$), the same baseline current may fall within the window, facilitating them. These predictions are consistent with the experimental findings of \cite{derjean_dynamic_2003}.

The absolute size of the bistability window matters equally for observing long-lasting afterdischarges, but it has not been directly quantified experimentally. However, studies using rats typically applied current pulses of 20 to $\qty{150}{\pico\ampere}$ \cite{derjean_dynamic_2003,morisset_ionic_1999,monteiro_switching-and_2006,reali_intrinsic_2011}, suggesting that the bistability window in these preparations falls within this range. Based on the estimated soma diameter of deep dorsal horn projection neurons of $\qty{20}{\micro\meter}$ \cite{derjean_dynamic_2003}, this corresponds to approximately $1.6$ to $\qty{11.9}{\micro\ampere\per\centi\meter^2}$, consistent with the range over which we defined our robust-bistability criterion in the minimal model. The upper bound remains less constrained and would require direct experimental measurements of $I_1$ and $I_2$. The resting states and firing frequencies observed experimentally could provide additional information to estimate the physiological range of bistability window sizes. However, these properties depend strongly on the full set of ionic conductances present in the neuronal membrane, making it difficult to infer physiological window sizes from a conductance-based model alone.

The firing frequencies in the bistability window of the minimal model (100 to $\qty{200}{\hertz}$) exceed those reported experimentally for deep projection neurons (5 to $\qty{20}{\hertz}$) \cite{morisset_ionic_1999,derjean_dynamic_2003,monteiro_switching-and_2006,reali_integrated_2005,reali_intrinsic_2011}. 
To assess whether this discrepancy limits the applicability of our results, we repeated the analysis of bistability in a two-compartment model of deep projection neurons incorporating additional ion channels \cite{le_franc_multiple_2010}, which produced firing frequencies more consistent with the experimentally reported range of $5$--$\qty{20}{\hertz}$~\cite{zhang_modeling_2014}. This supports the generalizability of our findings to more comprehensive models and their robustness to the presence of additional conductances.  The other ion channels may further modulate the bistability window. CAN channels have been linked to afterdischarge maintenance~\cite{morisset_ionic_1999,fossat_ltype_2007,le_franc_multiple_2010,aguiar_nmda_2010} and may therefore also promote bistability. KCa channels, which contribute to afterdischarge termination~\cite{morisset_ionic_1999} and oppose CaL channel effects on excitability \cite{drion_dynamic_2015,franci_balance_2013}, may reduce the bistability window and could account for the termination of persistent firing. The faster CaL subtype ($\mathrm{CaL,f}$) enlarged the window much less than the slower subtype in the minimal model (\nameref{app:CaLf}), suggesting that channel timescale matters as much as channel type. Finally, both CaL and CAN channels have been associated with windup \cite{aguiar_nmda_2010,le_franc_multiple_2010,fossat_ltype_2007,aby_windup_2019}, a form of short-term sensitization in projection neurons. Whether windup shares a mechanistic basis with bistability remains an open question.

In the dorsal horn, Kir channels are modulated by several signaling cascades, including GABA binding to GABA$_\mathrm{B}$ receptors, neuropeptide Y (NPY) binding to its Y1 receptor, and gastrin-releasing peptide (GRP) binding to its receptor (GRPR)~\cite{wood_potassium_2019,sinha_fast_2021,pagani_how_2019}. Bistability may therefore be a more general feature of dorsal horn circuitry, not restricted to projection neurons but potentially extending to interneuron populations that co-express Kir and calcium channels and are targeted by these signaling pathways. In this context, the persistent depolarization and spontaneous activity observed for minutes after synaptic stimulation in GRPR-expressing interneurons \cite{pagani_how_2019} may reflect underlying bistability. More broadly, Kir-supported bistability at the single-cell level may be useful wherever a neuron must retain a short-term memory of past input, as noted in thalamocortical and other systems~\cite{sanders_nmda_2013,amarillo_inward_2018,delmoe_conditions_2023}.

In summary, a minimal conductance-based model shows that CaL and Kir channels cooperate to produce robust and physiological bistability, plateau potentials, and sustained afterdischarges in deep dorsal horn projection neurons. The mechanism relies on a shape feature shared by both steady-state currents, namely a region of negative differential conductance around the spike threshold, which enlarges the bistability window without compromising the physiological plausibility of resting states. Because Kir channels are regulated by multiple neuromodulatory inputs in the dorsal horn, the CaL+Kir pathway is a plausible intrinsic target by which central sensitization switches the functional state of nociception.

\section*{Materials and methods}
\subsection*{Software}
    All simulations and experiments were performed in the Julia programming language. The code and packages used in this work are available on \href{https://github.com/anadew2/cal-kir-bistability}{GitHub}.

\subsection*{Minimal conductance-based model} 
    We used a single-compartment Hodgkin-Huxley model, in which the membrane potential $V$ evolves in response to an applied current $I$ according to:
    \begin{linenomath}
    \begin{eqnarray}
        C \frac{dV}{dt} = - I_\mathrm{leak} - \sum_{\mathrm{ion} \in \mathcal{I}} I_{\mathrm{ion}} + I , \label{eq:dV}
    \end{eqnarray}
    \end{linenomath}
    where $C$ is the membrane capacitance, $ I_{\mathrm{leak}}$ is the leak current, $I_{\mathrm{ion}}$ are the intrinsic ionic currents with $\mathcal{I}$ the set of all ionic channels.     
    In this work, the ionic currents flowing in each type of voltage-dependent ion channels are defined by:
    \begin{linenomath}
    \begin{eqnarray}
        I_{\mathrm{ion}} = \bar{g}_\mathrm{ion} \, m_{\mathrm{ion}}^{q_{\mathrm{ion}}} \, h_{\mathrm{ion}}^{r_{\mathrm{ion}}} \, (V-E_{\mathrm{ion}}) ,
    \end{eqnarray}
    \end{linenomath}
    where $\bar{g}_{\mathrm{ion}}$ denotes the ion channels maximal conductance, $m_{\mathrm{ion}}$ and $h_{\mathrm{ion}}$ correspond to the voltage-dependent activation and inactivation gates of the ion channels, $q_{\mathrm{ion}}$ is an integer between $1$ and $4$, $r_{\mathrm{ion}}$ is an integer between $0$ and $1$, and $E_{\mathrm{ion}}$ denotes the reversal potential of the considered ionic current. 

    The L-type calcium channels, embedding a calcium-dependency, were modeled using a Goldman–Hodgkin–Katz formalism \cite{hodgkin_effect_1949} such that: 
    \begin{linenomath}
    \begin{eqnarray}
        I_{\mathrm{CaL}} = \bar{p}_\mathrm{CaL} \, m_{\mathrm{CaL}}^{2} \, h_{\mathrm{CaL}}\mathrm{GHK}(V,\left[\mathrm{Ca}^{2+} \right]_i),
    \end{eqnarray}
    \end{linenomath}
    with
    \begin{linenomath}
    \begin{eqnarray}
        \mathrm{GHK}(V,\left[\mathrm{Ca}^{2+} \right]_i) =  10^{-3} \cdot 2F \cdot \left( \left[\mathrm{Ca}^{2+} \right]_i \cdot \dfrac{-w}{e^{-w} -1} - \left[\mathrm{Ca}^{2+} \right]_o \cdot \dfrac{w}{e^{w} -1}\right) ,
    \end{eqnarray}
    \end{linenomath}
    and
    \begin{linenomath}
    \begin{eqnarray}
        w = 10^{-3}\cdot V\cdot \dfrac{2F}{RT},
    \end{eqnarray}
    \end{linenomath}
    where $F$ is the Faraday constant, $R$ is the perfect gas constant, $T$ is the temperature, $\left[\mathrm{Ca}^{2+} \right]_i $ is the intracellular calcium concentration, and $\left[\mathrm{Ca}^{2+} \right]_o $ is the extracellular calcium concentration, considered constant. 

    The evolution of the intracellular calcium concentration was modeled as: 
    \begin{linenomath}
    \begin{eqnarray}\label{eq:dCadt}
        \dfrac{d\left[\mathrm{Ca}^{2+} \right]_i}{dt} = - 10^7 \cdot \dfrac{I_{\mathrm{CaL}}}{10^3} \cdot \dfrac{1}{2Fd} + \dfrac{\left(\left[\mathrm{Ca}^{2+} \right]_{i,0}-\left[\mathrm{Ca}^{2+} \right]_i \right) }{\tau_\mathrm{Ca}},
    \end{eqnarray}
    \end{linenomath}
    where $I_{\mathrm{CaL}}$ is the intrinsic ionic current flowing in the L-type calcium channels $F$ is the Faraday constant, $d$ is the depth of the shell beneath the membrane, $\left[\mathrm{Ca}^{2+} \right]_{i,0}$ is the initial intracellular calcium concentration, and $\tau_\mathrm{Ca}$ is the buffering time constant. 
    
    The evolutions of the activation and inactivation gates of a given ion channel are defined as:
    \begin{linenomath}
    \begin{align*}
        \frac{dm_{\mathrm{ion}}}{dt} &= \left(m_{\mathrm{ion},\infty}(V)-m_{\mathrm{ion}} \right)/ \tau_{m_{\mathrm{ion}}}(V) , \\
        \frac{dh_{\mathrm{ion}}}{dt} & = \left(h_{\mathrm{ion},\infty}(V)-h_{\mathrm{ion}} \right)/ \tau_{h_{\mathrm{ion}}}(V) , 
    \end{align*}
    \end{linenomath}
    where $m_{\mathrm{ion},\infty}(V)$ and $h_{\mathrm{ion},\infty}(V)$ denote the steady-state values of the activation and inactivation gates at a given membrane potential $V$, and $\tau_{m_{\mathrm{ion}}}(V)$ and $\tau_{h_{\mathrm{ion}}}(V)$ the time constants of the activation and inactivation gates at a given membrane potential $V$, respectively.

    The minimal model includes a voltage-gated fast sodium current $I_{\mathrm{Na}}$, a slow delayed-rectifier potassium current $I_{\mathrm{KDR}}$, an L-type calcium current $I_{\mathrm{CaL}}$, an M-type potassium current $I_{\mathrm{KM}}$, an inward rectifier potassium current $I_{\mathrm{Kir}}$, and a leak current $I_{\mathrm{leak}}$. The models of $I_{\mathrm{Na}}$ and $I_{\mathrm{KDR}}$ follow~\cite{destexhe_model_1994}. The model of $I_{\mathrm{CaL}}$ corresponds to the slower L-type calcium current of~\cite{le_franc_multiple_2010}, with its steady-state activation slightly modified to follow a Boltzmann equation; the faster L-type calcium current, used only in the Appendix, is described there. The models of $I_{\mathrm{Kir}}$ and $I_{\mathrm{leak}}$ are also from~\cite{le_franc_multiple_2010}. The model of $I_{\mathrm{KM}}$ follows~\cite{miceli_early-onset_2015}, with the half-activation voltage lowered to \SI{-76}{\milli\volt} and the slope factor increased to \SI{25.7}{\milli\volt}. These modifications bring the activation range of KM closer to that of Kir, to better contrast their effects. An unmodified version is used in the Appendix and yields the same conclusions (see~\nameref{app:other_KM}).

    Throughout, ionic currents are expressed in \si{\micro\ampere\per\centi\meter^2}, maximal conductances in \si{\milli\siemens\per\centi\meter^2}, the L-type calcium permeability in \si{\centi\meter\per\second}, calcium concentrations in \si{\milli\molar}, the membrane potential in \si{\milli\volt}, and time in \si{\milli\second}. Time constants, steady-state activations, and inactivations for $I_{\mathrm{CaL}}$, $I_{\mathrm{KM}}$, and $I_{\mathrm{Kir}}$ are given in Table~\ref{tab:minfandtau}. The nominal values of the fixed parameters are given in Tables~\ref{tab:fixedparam1} and~\ref{tab:fixedparam2}.

    \begin{table}[!ht]
        \begin{adjustwidth}{-2.25in}{0in} 
        \centering
        \caption{
        {\bf Steady-state activations and inactivation used to model $I_{\mathrm{CaL}}$, $I_{\mathrm{Kir}}$, and $I_{\mathrm{KM}}$.}}
        \begin{tabular}{|c|c|c|c|c|}
        \hline
        $x_{\mathrm{ion}}$ & $x_{\mathrm{ion},\infty}(V)$ & $\tau_{x_{\mathrm{ion}}}(V)$ & $q_{\mathrm{ion}}$ & $r_{\mathrm{ion}}$ \\ \thickhline
        $m_{\mathrm{CaL}}$ & $\dfrac{1}{1 + \exp(-(V+\num{25.4})/6)}$ & $\dfrac{160}{\frac{0.1(-V-40)}{\exp{(0.1(-V-40))} -1 + 4 \exp{((-V-65)/18)}}}$ & / & / \\ \hline
        $h_{\mathrm{CaL}}$ & $\dfrac{1}{1 + \exp((V+14)/\num{4.03})}$ & $10000$ & / & / \\ \hline
        $m_{\mathrm{Kir}}$ & $\dfrac{1}{1 + \exp((V+65)/10)}$ & $1$ & 1 & 0 \\ \hline
        $m_{\mathrm{KM}}$ & $\dfrac{1}{1 + \exp((-V-\num{76,1})/\num{25,7})}$ & $103$ & 1 & 0 \\ \hline
        \end{tabular}
        \begin{flushleft} 
        \end{flushleft}
        \label{tab:minfandtau}
        \end{adjustwidth}
    \end{table}

    \begin{table}[!ht]
        \begin{adjustwidth}{-2.25in}{0in}
        \centering
        \caption{
        {\bf Nominal values of the fixed parameters of the conductance-based model.}}
        \begin{tabular}{|c|c|c|c|c|c|c|c|}
        \hline
        $C$ & $E_{\mathrm{Na}}$ & $E_{\mathrm{K}}$ &  $E_{\mathrm{leak}}$ & $\bar{g}_{\mathrm{Na}}$ & $\bar{g}_{\mathrm{KDR}}$ & $g_{\mathrm{leak}}$  \\ \thickhline
        \SI{1}{\micro \farad / \centi\meter^2} & \SI{50}{\milli \volt} & \SI{-90}{\milli \volt} &  \SI{-60.1}{\milli \volt} & \SI{30}{\milli \siemens / \centi\meter^2} & \SI{4}{\milli \siemens / \centi\meter^2} & \SI{0.033}{\milli \siemens / \centi\meter^2} \\ \hline
        \end{tabular}
        \begin{flushleft} 
        \end{flushleft}
        \label{tab:fixedparam1}
        \end{adjustwidth}
    \end{table}
    
    \begin{table}[!ht]
        \begin{adjustwidth}{-2.25in}{0in} 
        \centering
        \caption{
        {\bf Nominal values of the fixed parameters of the conductance-based model.}}
        \begin{tabular}{|c|c|c|c|c|c|c|c|}
        \hline
        $F$ & $R$ & $T$ & $d$  & $\tau_\mathrm{Ca}$ & $\left[\mathrm{Ca}^{2+} \right]_o $ & $\left[\mathrm{Ca}^{2+} \right]_{i,0}$ \\ \thickhline
        \SI{96520}{\coulomb\per\mole} & \SI{8.31}{\joule\per\mole\kelvin} & \SI{309.15}{\kelvin} & \SI{1}{\nano\meter} & \SI{10}{\milli\second} &  \qty{2}{\milli\molar} & \qty{5e-5}{\milli\molar}  \\ \hline
        \end{tabular}
        \begin{flushleft} 
        \end{flushleft}
        \label{tab:fixedparam2}
        \end{adjustwidth}
    \end{table}

\subsection*{Complete conductance-based model}

    To compute the results shown in Fig.~\ref{fig:comp}, we reimplemented a published two-compartment conductance-based model of deep projection neurons (see\cite{le_franc_multiple_2010} for more details). The ion currents and the fixed parameters described in the previous section are identical to those used in the two-compartment model. This model includes two types of L-type voltage-gated calcium channels as described in\cite{le_franc_multiple_2010}. The CaL current described in the previous section corresponds to the slower type, denoted "CaLs" in this paper. In addition, the two-compartment model includes calcium-dependent potassium (KCa) channels and calcium-dependent nonspecific cation (CAN) channels.

\subsection*{Current step and current pulse simulations} 
    A time-dependent applied current was used to observe the neuron response to a current step and pulse. To create a step of current, the time dependency was introduced such that the applied current $I$ is equal to:        
    \begin{linenomath}
    \begin{eqnarray}
        I(t) = I_{s0} + (I_{s1} - I_{s0}) \unitstep(t - t_{\text{step}}),
    \end{eqnarray}
    \end{linenomath}
    where $\unitstep(x)$ is the unit step function such that it equals $1$ if $x \geq 0$ and $0$ if $x < 0$.   
    This current step was depolarizing when $I_{s1}>I_{s0}$ and hyperpolarizing otherwise. 

    To create a pulse of current, the time dependency was introduced such that the applied current $I$ is equal to:
    \begin{linenomath}
    \begin{eqnarray}
            I(t) = I_{p0} + (I_{p1} - I_{p0}) [ \unitstep(t - t_{\text{p,starts}}) - \unitstep(t - t_{\text{p,ends}}) ] ,
    \end{eqnarray}
    \end{linenomath}
    meaning that this current pulse was depolarizing when $I_{p1}>I_{p0}$ and hyperpolarizing otherwise. 

   Both protocols probe bistability. In the step protocol, ascending steps from rest and descending steps from spiking to the same final currents reveal the bistability window by showing which currents lead to different final states depending on initial conditions (Figs~\ref{fig:1}A--B). In the pulse protocol, a depolarizing pulse with baseline $I_1<I_{p0} < I_2$ and amplitude above $I_2$ triggers a transition to spiking that terminates when the pulse ends, tracing the same window boundaries (Figs~\ref{fig:2}A,D). We use the step protocol for parameter sweeps and the pulse protocol to illustrate afterdischarges. 

    In Fig.~\ref{fig:1}A, we applied four hyperpolarizing steps of current at $t_{\text{step}} = \SI{250}{\milli\second}$ with $I_{s0} =$\SI{-3}{\micro\ampere/ \centi\meter ^2} and $I_{s1}$ taking one of the four values in the ensemble $\{-2.5;-1.5;-0.5;0.5\}$ \SI{}{\micro\ampere/ \centi\meter ^2} from bottom to top. In Fig.~\ref{fig:1}B, we applied four depolarizing steps of current at $t_{\text{step}} = \SI{250}{\milli\second}$ with $I_{s0} =$\SI{1}{\micro\ampere/ \centi\meter ^2} and $I_{s1}$ taking one of the four values in the ensemble $\{-2.5;-1.5;-0.5;0.5\}$ \SI{}{\micro\ampere/ \centi\meter ^2} from bottom to top. For Figs.~\ref{fig:1}A and \ref{fig:1}B, we used $\bar{p}_{\mathrm{CaL}} = \qty{1.5e-5}{\centi\meter\per\second}$ and $\bar{g}_{\mathrm{KM}}=\bar{g}_{\mathrm{Kir}}=0$.  

     In Fig.~\ref{fig:2}A, we applied a depolarizing pulse of current from $t_{\text{p,starts}} = \SI{250}{\milli\second}$ to $t_{\text{p,ends}} = \SI{350}{\milli\second}$ with $I_{s0} =$\SI{1}{\micro\ampere/ \centi\meter ^2} and $I_{s1} =$ \SI{4}{\micro\ampere/ \centi\meter ^2} to the model defined by the set of maximal conductances: $\bar{p}_{\mathrm{CaL}} = \qty{1.5e-5}{\centi\meter\per\second}$, $\bar{g}_{\mathrm{KM}} = \SI{0.15}{\milli\siemens\per\centi\meter^2}$, and $\bar{g}_{\mathrm{Kir}} = \SI{0}{\milli\siemens\per\centi\meter^2}$. 
     In Fig.~\ref{fig:2}D, the same depolarizing pulse of current was adapted such that $I_{p0} =$\SI{-0.8}{\micro\ampere/ \centi\meter ^2} and $I_{p1} =$ \SI{2.2}{\micro\ampere/ \centi\meter ^2}  to the model defined by the set of maximal conductances: $\bar{p}_{\mathrm{CaL}} = \qty{1.5e-5}{\centi\meter\per\second}$, $\bar{g}_{\mathrm{KM}} = \SI{0}{\milli\siemens\per\centi\meter^2}$, and $\bar{g}_{\mathrm{Kir}} = \SI{0.1}{\milli\siemens\per\centi\meter^2}$. 

    In Fig.~\ref{fig:comp}A-D, we applied several depolarizing pulses of current from $t_{\text{p,starts}}=\qty{500}{\milli\second}$ to $t_{\text{p,ends}} = \qty{2000}{\milli\second}$. The values of $I_{p0}$, $I_{p1}$, $\bar{p}_{\mathrm{CaL}}$ and $\bar{g}_{\mathrm{Kir}}$ used are given in the bottom part of each of these figures.

    In Fig.~\ref{fig:8}, a hyperpolarizing step was applied at $t_{\text{step}} = \SI{4}{\second}$ around $I_1$, with $I_{s0} = I_1 + \SI{10}{\nano\ampere/\centi\meter^2}$ and $I_{s1} = I_1 - \SI{10}{\nano\ampere/\centi\meter^2}$ for each set of conductances. Panel A used $\{I_{s0};I_{s1}\} = \{-0.05;-0.07\}~\si{\micro\ampere/\centi\meter^2}$ and $\{\bar{p}_{\mathrm{CaL}};\bar{g}_{\mathrm{KM}};\bar{g}_{\mathrm{Kir}}\} = \{0.;0.;0.\}$; panel B used $\{-0.42;-0.44\}$ and $\{\qty{7.5e-6}{};0.;0.\}$; panel C used $\{3.6;3.58\}$ and $\{0.;0.;0.3\}$\, ; panel D used $\{6.6;6.58\}$ and $\{0.;0.3;0.\}$. The same step protocol was applied in Fig.~\ref{fig:9-CaLs+Kir/KM}A,B: panel A used $\{2.86;2.84\}$ and $\{\qty{7.5e-6};0.;0.3\}$; panel B used $\{6.05;6.03\}$ and $\{\qty{7.5e-6};0.3;0.\}$.

\subsection*{Computation of the frequency-current curves}       
    A frequency-current (fI) curve represents the steady-state spiking frequency observed at a constant applied current. In the context of a bistable neuron, an overlap exists between a zero-frequency state (\ie, a resting state) and a non-zero-frequency state (\ie, a spiking state) within a range of current values in which bistability is observed. In this work, we designate this range of current the \textit{bistability window}, which is defined between the onset of spiking at current $I_1$ and the cessation of resting at $I_2$. 
    
    Thus, the initial conditions used when computing the steady-state spiking frequency for currents within the bistability window must allow the neuron to converge to its spiking state. To determine them, we initialized the neuron at resting and applied a depolarizing current step from the $i^{\text{th}}$ current (in the range of current tested) to a higher current, out of the bistability window. 
    Then, the final state observed was used as the initial conditions to simulate the $i^{\text{th}}$ constant current and determine the corresponding steady-state spiking frequency. This ensures that the ultra-slow inactivation gate of calcium channels is still open, and facilitates the convergence toward spiking, even with low steady-state frequencies. The model is then simulated for windows of \SI{120}{\second} until the spiking frequency reaches steady-state. 
    To compute the line of zero-frequency of the fI curve, we computed the model fixed points and incorporated a zero-frequency point if a stable fixed point was identified. The combination of these 2 computations was used in Fig.~\ref{fig:1}C (top) and Figs.~\ref{fig:2}B and E. In Fig.~\ref{fig:1}C (top), we used the maximal conductances: $\bar{p}_{\mathrm{CaL}} = \qty{1.5e-5}{\centi\meter\per\second}$ and $\bar{g}_{\mathrm{KM}} = \bar{g}_{\mathrm{Kir}} = \SI{0}{\milli\siemens\per\centi\meter^2}$, and a range of applied current between $-2.65$ and $\qty{1}{\micro\ampere\per\centi\meter^2}$, with a total of $129$ points. In Figs.~\ref{fig:2}B, the maximal conductance $\bar{g}_{\mathrm{KM}}$ was increased to $\SI{0.15}{\milli\siemens\per\centi\meter^2}$, with $\bar{p}_{\mathrm{CaL}} = \qty{1.5e-5}{\centi\meter\per\second}$ and $\bar{g}_{\mathrm{Kir}} = \SI{0}{\milli\siemens\per\centi\meter^2}$, and a range of applied current between $0.7$ and $\qty{4}{\micro\ampere\per\centi\meter^2}$, with a total of $125$ points. Similarly, the conductance $\bar{g}_{\mathrm{Kir}}$ was increased to $\SI{0.15}{\milli\siemens\per\centi\meter^2}$ in Fig.~\ref{fig:2}E, with $\bar{p}_{\mathrm{CaL}} = \qty{1.5e-5}{\centi\meter\per\second}$ and $\bar{g}_{\mathrm{KM}} = \SI{0}{\milli\siemens\per\centi\meter^2}$, and a range of applied current between $-1.1$ and $\qty{2.4}{\micro\ampere\per\centi\meter^2}$, with a total of $127$ points.

    In Fig.~\ref{fig:altgt}C, the minimal model was used with $\bar{p}_{\mathrm{CaL}}=\qty{1.5e-5}{\centi\meter\per\second}$ and three $\{\bar{g}_{\mathrm{KM}};\bar{g}_{\mathrm{Kir}}\}$ pairs (in \si{\milli\siemens\per\centi\meter^2}): $\{0.02;0.28\}$ (pink, $0.6$--$\qty{5.6}{\micro\ampere\per\centi\meter^2}$, 133 points), $\{0.28;0.02\}$ (purple, $3.9$--$\qty{9.9}{}$, 132 points), and $\{0.28;0.28\}$ (yellow, $6.6$--$\qty{9.8}{}$, 120 points).

       In Fig.~\ref{fig:comp}E,F (top), the complete model was used with four $\{\bar{p}_{\mathrm{CaL}};\bar{g}_{\mathrm{Kir}}\}$ pairs: $\{\qty{2.0e-6}{};0.015\}$ (blue, $5$--$\qty{250}{\pico\ampere}$, 1030 points), $\{\qty{2.0e-6}{};0.185\}$ (pink, $161$--$250$, 893 points), $\{\qty{1.5e-5}{};0.015\}$ (orange, $-134$--$149$, 1038 points), and $\{\qty{1.5e-5}{};0.185\}$ (green, $14$--$247$, 1028 points).

    In Fig.~\ref{fig:1}C (bottom), and Figs.~\ref{fig:2}C and F, we also represented the resting equilibria observed for the range of current used to display the fI curve. The resting equilibria were computed together with the line of zero-frequency, by computing the membrane potential associated with the stable fixed point identified for an input range of applied current. The maximal conductances used in Figs.~\ref{fig:1}C (bottom), \ref{fig:2}C and \ref{fig:2}F were identical to the maximal conductances used in Figs.~\ref{fig:1}C (top), \ref{fig:2}B and \ref{fig:2}E, respectively. Because the computation of the resting equilibria was much faster than the computation of the fI curve, we were able to use an even more refined range of current for these figures. Indeed, we used a range of applied current between $-3$ and $\qty{-0.1}{\micro\ampere\per\centi\meter^2}$ with a total of $8914$ points for Fig.~\ref{fig:1}C (bottom), between $0$ and $\qty{3}{\micro\ampere\per\centi\meter^2}$ with a total if $2273$ points for Fig.\ref{fig:2}C, and between $-1.6$ and $\qty{1.7}{\micro\ampere\per\centi\meter^2}$ with a total of $2920$ points for Fig.\ref{fig:2}E. Note that there were no stable fixed points existing above these ranges of current. 

    In Fig.~\ref{fig:altgt}C(bottom), the curves defining the resting equilibrium observed for a given current were computed on ranges of current between $0$ and $\qty{3.6}{\micro\ampere\per\centi\meter^2}$ with a total of $2207$ points for the pink curve,  between $0$ and $\qty{6.2}{\micro\ampere\per\centi\meter^2}$ with a total of $2278$ points for the purple curve, and between $0$ and $\qty{8.8}{\micro\ampere\per\centi\meter^2}$ with a total of $2989$ points for the yellow curve. The set of conductances used are identical to those used in Fig.~\ref{fig:altgt}C(top) for each color. 

    In Fig.~\ref{fig:comp}E-F(bottom), the curves defining the resting equilibrium observed for a given current were computed on ranges of current between $-150$ and $\qty{55}{\pico\ampere}$ with a total of $492$ points for the blue curve, between $-150$ and $\qty{212}{\pico\ampere}$ with a total of $284$ points for the pink curve, between $-150$ and $\qty{212}{\pico\ampere}$ with a total of $283$ points for the green curve, and between $-150$ and $\qty{49}{\pico\ampere}$ with a total of $480$ points for the orange curve. The set of conductances used are identical to those used in Fig.~\ref{fig:comp}E-F(top) for each color.

\subsection*{Computation of the bistability window heatmaps}          
    The computation of the fI curve is associated with a given set of conductances. 
    To capture the evolution of the bistability window size, we varied the pair of conductances, $\bar{p}_\mathrm{CaL}$ and either $\bar{g}_\mathrm{KM}$ or $\bar{g}_\mathrm{Kir}$, and computed for each pair of conductances the (absolute) size of the bistability window, noted: 
    \begin{linenomath}
    \begin{eqnarray}
        \Delta I = I_2-I_1.
    \end{eqnarray}
    \end{linenomath}
    The current $I_2$, which marks the end of the bistability window, was determined by identifying the highest current with a stable resting state. The stable resting states were computed using the same methodology to compute the fI curve zero-frequency line. 
    
    In contrast to the methodology employed in the computation of the descending fI curve, which determined the steady-state spiking frequency for each current within a specified range, in this instance, only the initial current, $I_1$, situated at the beginning of the bistability window, was subjected to analysis. To compute $I_1$ for a given pair of conductances in a reasonable execution time, this algorithm was modified by employing an ascending range of current and initiated at a current from the resting-only regime. In this manner, the current $I_1$ was still identified as the lowest current at which spiking with a constant frequency can be observed over time intervals of \SI{120}{s}, and with the same methodology employed to compute the neuron initial conditions (extracted from the end of a pulse simulation).
    The results of this analysis are shown in Fig.~\ref{fig:3}A. In Fig.~\ref{fig:3}D, we performed the same experiment by varying the maximal conductance $\bar{g}_\mathrm{Kir}$ instead of $\bar{g}_\mathrm{KM}$.

    Similarly, the resting equilibrium at the beginning of the bistability window ($\overline{V}(I_1)$) was determined as the membrane potential of the stable fixed point exhibited at $I_1$, following the variation of either pair of conductances $(\bar{p}_{\mathrm{CaL}}, \bar{g}_\mathrm{KM})$ or $(\bar{p}_{\mathrm{CaL}}, \bar{g}_\mathrm{Kir})$.
    This methodology was used for Fig.~\ref{fig:3}C and Fig.~\ref{fig:3}F, respectively. 

    The heatmaps shown in Fig.~\ref{fig:altgt}A-B and Fig.~\ref{fig:comp}G-I were computed in the same manner as what is described here above, with the minimal model and the complete model, respectively. The conductances $\bar{g}_{\mathrm{KM}}$ and $\bar{g}_{\mathrm{Kir}}$ were varied in ranges each defined between $0$ and $\qty{0.3}{\milli\siemens\per\centi\meter^2}$ including 31 points (Fig.~\ref{fig:altgt}A-B). The conductances $\bar{p}_{\mathrm{CaL}}$ and $\bar{g}_{\mathrm{Kir}}$ were varied in ranges defined between $0$ and $\qty{3e-5}{r\centi\meter\per\second}$ (with 41 points) and $0$ and $\qty{0.3}{\milli\siemens\per\centi\meter^2}$ (with 31 points), respectively (Fig.~\ref{fig:comp}G-I).
    
\subsection*{Response to noisy input currents}

    To evaluate the robustness of resting and spiking within the bistability window, we applied a constant current centered on the window of each pair of conductances and added white noise to the voltage equation.
    For this purpose, we simulated the ODE described in \ref{eq:dV} with additive white noise, following a Normal distribution with zero-mean and a standard deviation of $\sigma$. 
    The parameter $\sigma$ was varied between \SI{0}{} and \SI{9}{\milli\volt} to modify the noise amplitude. The system of equations formed by this stochastic differential equation and the ordinary differential equation associated with the evolution of the ion channels gates and the intracellular calcium concentration was solved with a fixed time step of \SI{0.01}{\milli\second}.

    In Figs.~\ref{fig:4}A, \ref{fig:4}B, \ref{fig:4}D and \ref{fig:4}E, $\sigma$ was increased at each time step, for a total simulation time of \SI{5}{s}. In Figs.~\ref{fig:4}A and \ref{fig:4}D, the set of initial conditions of the experiment was the resting state observed in the center of the bistability window of the pair of conductances considered (either $(\bar{p}_{\mathrm{CaL}},\bar{g}_\mathrm{KM})$ or $(\bar{p}_{\mathrm{CaL}}, \bar{g}_\mathrm{Kir})$, respectively). Reciprocally, in Figs.~\ref{fig:4}B and \ref{fig:4}E, the set of initial conditions of the experiment was a point of the spike trajectory obtained after applying the center of the bistability window of the pair of conductances considered (either $(\bar{p}_{\mathrm{CaL}},\bar{g}_\mathrm{KM})$ or $(\bar{p}_{\mathrm{CaL}}, \bar{g}_\mathrm{Kir})$, respectively), without noise for \SI{80}{\second}, such that spiking has reached steady-state before applying any noise.

    In Figs.~\ref{fig:4}C and \ref{fig:4}F, we evaluated the global neuron ability to sustain either a resting or a spiking state in the presence of a fixed noise amplitude for each type of bistability, which was produced by coupling either $(\bar{p}_{\mathrm{CaL}},\bar{g}_\mathrm{KM})$ or $(\bar{p}_{\mathrm{CaL}}, \bar{g}_\mathrm{Kir})$, respectively. To that end, the stochastic model was simulated for a period of \SI{50}{\second} with a constant value of $\sigma$ and initialized to either the resting or the spiking state, both computed from a \SI{80}{\second} noise-free simulation. If the neuron behavior, initially at resting (resp. spiking) transitioned to spiking (resp.\ resting) at the end of the simulation, a value of 1 was stored as a marker of this state transition, and 0 otherwise. 
    This experiment was repeated 500 times for each standard deviation value tested. The ability to sustain the initial behavioral pattern was quantified by the \textit{proportion of state transition} and calculated as the mean value of the state transition markers recorded for each of the 500 simulations.
    Both Figs.~\ref{fig:4}C and \ref{fig:4}F display the proportion of state transition given the constant value of $\sigma$ when the neuron is initially spiking or resting. In Fig.~\ref{fig:4}C, bistability relies on the pair $(\bar{p}_{\mathrm{CaL}},\bar{g}_\mathrm{KM})$ while in Fig.~\ref{fig:4}F, bistability relies on the pair $(\bar{p}_{\mathrm{CaL}}, \bar{g}_\mathrm{Kir})$.

    The two sets of conductances were matched for window size. Fig.~\ref{fig:4}A,B,C used $\{\bar{p}_{\mathrm{CaL}};\bar{g}_{\mathrm{KM}};\bar{g}_{\mathrm{Kir}}\} = \{\qty{1.5e-5}{};0.3;0\}$, giving $\Delta I = \SI{1.77}{\micro\ampere/\centi\meter^2}$; Fig.~\ref{fig:4}D,E,F used $\{\qty{1.16e-5}{};0;0.3\}$, giving $\Delta I = \SI{1.73}{\micro\ampere/\centi\meter^2}$.

\subsection*{Assessment of the reachability of the resting and spiking states}
    To complement our analysis of the robustness of the resting and spiking states within the bistability window for each pair of conductances considered, we assessed the reachability of each state. For that purpose, we simulated the system of ODE described in \ref{eq:dV} with a constant value of applied current $I$ sampled from the range $[I_1;I_2]$, delimiting each bistability window obtained with $(\bar{p}_{\mathrm{CaL}},\bar{g}_\mathrm{KM})$ or $(\bar{p}_{\mathrm{CaL}}, \bar{g}_\mathrm{Kir})$. For each value of $I$ tested, we used a range of initial membrane potential $V_0$ to define different sets of initial conditions written as: 
    $$(V_0,m_{\mathrm{ion},\infty}(V_0),h_{\mathrm{ion},\infty}(V_0), ... , \left[\mathrm{Ca}^{2+} \right]_0), \hspace{1cm} \text{with } V_0 \in [-90;-20] \qty{ }{\milli\volt}, $$
    where each activation and inactivation gate were initialized to the value of their steady-state function in $V_0$ (\ie $m_{\mathrm{ion},\infty}(V_0)$ and $h_{\mathrm{ion},\infty}(V_0)$, respectively). The initial calcium concentration $\left[\mathrm{Ca}^{2+} \right]_0$ was set to \qty{5e-5}{\milli\molar}. For each set of initial conditions and current, we determined the state toward which the neuron converged to, which was either resting or spiking.  

     In Fig.~\ref{fig:6B}A, the following conductances were used: $\bar{p}_{\mathrm{CaL}} = \qty{1.16e-5}{\centi\meter\per\second}$, $\bar{g}_{\mathrm{KM}} = \SI{0}{\milli\siemens\per\centi\meter^2}$, and $\bar{g}_{\mathrm{Kir}} = \SI{0.3}{\milli\siemens\per\centi\meter^2}$, which created a bistability window of $\Delta I = \SI{1.73}{\micro\ampere/ \centi\meter ^2}$ with $I_1 = \qty{1.83}{\micro\ampere\per\centi\meter^2}$ and $I_2 = \qty{3.56}{\micro\ampere\per\centi\meter^2}$. In Fig.~\ref{fig:6B}B, the following set of conductances was used $\bar{p}_{\mathrm{CaL}} = \qty{1.5e-5}{\centi\meter\per\second}$, $\bar{g}_{\mathrm{KM}} = \SI{0.3}{\milli\siemens\per\centi\meter^2}$, and $\bar{g}_{\mathrm{Kir}} = \SI{0}{\milli\siemens\per\centi\meter^2}$, which created a bistability window of $\Delta I = \SI{1.77}{\micro\ampere/ \centi\meter ^2}$ with $I_1 = \qty{4.67}{\micro\ampere\per\centi\meter^2}$ and $I_2 = \qty{6.44}{\micro\ampere\per\centi\meter^2}$. For both Fig.~\ref{fig:6B}A and B, we used a range of current from $I_1$ to $I_2$ with a step of $\qty{0.035}{\micro\ampere\per\centi\meter^2}$ such that it contained 50 data points, and a range of $V_0$ defined between $\qty{-90}{\milli\volt}$ and $\qty{-20}{\milli\volt}$ with a step of $\qty{0.1}{\milli\volt}$ containing 701 data points. 

\subsection*{Robustness to different levels of intrinsic variability}    
    Each pair of conductances $(\bar{p}_{\mathrm{CaL}},\bar{g}_\mathrm{KM})$ or $(\bar{p}_{\mathrm{CaL}}, \bar{g}_\mathrm{Kir})$ give rise to two types of bistability. We tested their robustness by measuring the relative change in the bistability window size $\Delta I$ for each pair considering neurons with heterogeneous intrinsic membrane properties.
    
    To achieve this, three levels of intrinsic variability were introduced, comprising 10, 20, and \SI{30}{\percent} in three parameters: the membrane capacitance $C$, the voltage-gated sodium channels maximal conductance $\bar{g}_\mathrm{Na}$, and the leak conductance $g_\mathrm{leak}$. Each parameter was randomly chosen in an interval centered on the corresponding nominal value used in the model. More specifically, in the case of a level of \SI{10}{\percent} of variability, the membrane capacitance was sampled according to a uniform distribution from $C (1-0.1)$ to $C (1+0.1)$. The same procedure was repeated to sample $\bar{g}_\mathrm{Na}$ and $g_\mathrm{leak}$. Subsequently, we computed $I_1$ and $I_2$, the limits of the bistability window, using the same algorithm as for the bistability window heatmaps, and computed the relative change in the bistability window size $\Delta I$ to its nominal value. This computation was realized for both pairs $(\bar{p}_{\mathrm{CaL}},\bar{g}_\mathrm{KM})$ and $(\bar{p}_{\mathrm{CaL}}, \bar{g}_\mathrm{Kir})$. The results of the $\mathrm{KM}$ model (in purple tones) and the $\mathrm{Kir}$ model (in pink tones) in Fig.~\ref{fig:5} were obtained with the sets of conductances $\{\bar{p}_{\mathrm{CaL}};\bar{g}_{\mathrm{KM}};\bar{g}_{\mathrm{Kir}}\} = \{\qty{1.5e-5};0.3;0\} $ and $\{\bar{p}_{\mathrm{CaL}};\bar{g}_{\mathrm{KM}};\bar{g}_{\mathrm{Kir}}\} = \{\qty{1.5e-5};0;0.3\} $, respectively.

\subsection*{Computation of the steady-state currents and the differential conductances} 
    Fig.~\ref{fig:6} (top) illustrates the steady-state currents of $I_{\mathrm{CaL}}$, $I_{\mathrm{KM}}$, and $I_{\mathrm{Kir}}$. At a given membrane potential, these steady-state currents are defined by the current amplitude with the activation and inactivation gates at steady-state, that is:
    \begin{linenomath}
    \begin{eqnarray}
        I_{\mathrm{ion}, \infty}(V) = \bar{g}_\mathrm{ion}\left ( m_{\mathrm{ion},\infty}(V) \right )^{q_{\mathrm{ion}}} \left (h_{\mathrm{ion},\infty}(V) \right ) ^{r_{\mathrm{ion}}}(V-E_{\mathrm{ion}}),
    \end{eqnarray}
    \end{linenomath}
    for voltage-dependent ion channels, and 
    \begin{linenomath}
    \begin{eqnarray}I_{\mathrm{CaL},\infty}\left(V,\left[\mathrm{Ca}^{2+} \right]_i \right ) = \bar{p}_\mathrm{CaL} \, m_{\mathrm{CaL},\infty}^{2}(V) \, h_{\mathrm{CaL},\infty}(V) \ \mathrm{GHK}(V,\left[\mathrm{Ca}^{2+} \right]_i),
    \end{eqnarray}
    \end{linenomath}
    for the L-type calcium channels. 

    Based on this definition, the differential conductance shown in Fig.~\ref{fig:6} (bottom) is the derivative of the steady-state current to the membrane potential ($\partial I_{\mathrm{ion}, \infty} / \partial V$) (method inspired from \cite{drion_dynamic_2015}).
    Each derivative of the steady-state ionic currents were computed with an Euler method, and the intracellular calcium concentration was set to a constant. In Fig.~\ref{fig:6}, we used $\left[\mathrm{Ca}^{2+} \right]_i = \left[\mathrm{Ca}^{2+} \right]_o $.

\subsection*{Analysis of the contribution of inward and outward currents} 
    To investigate the reasons behind the differing impacts of KM and Kir channels on the bistability window size, we modified these channels separately to block their outward (resp.\ inward) currents. This approach allowed us to analyze the contribution of their inward (resp.\ outward) currents alone. The resulting inward currents flowing in these modified channels are designated by $I_{\mathrm{KM, inward}}$ and $I_{\mathrm{Kir, inward}}$, while the outward currents are denoted $I_{\mathrm{KM, outward}}$ and $I_{\mathrm{Kir, outward}}$. This modification is applied to the steady-state functions of the activation gates of both KM and Kir channels, both defined without an inactivation gate, realized as follows:
    \begin{linenomath}
    \begin{eqnarray}
        m_{\mathrm{ion}, \mathrm{inward}, \infty}(V) = 
        \begin{cases}
            m_{\mathrm{ion} ,\infty}(V), & \text{if $V\leq E_{\mathrm{K}}$} ,  \\
            0, & \text{otherwise} ,
       \end{cases}   \\
       \qquad \text{and} \qquad    
       m_{\mathrm{ion}, \mathrm{outward}, \infty}(V) = 
        \begin{cases}
            m_{\mathrm{ion}, \infty}(V), & \text{if $V> E_{\mathrm{K}}$} ,  \\
            0, & \text{otherwise} .
       \end{cases} 
    \end{eqnarray}
    \end{linenomath}

       Fig.~\ref{fig:7} (left) represents the steady-state inward and outward KM and Kir currents. To compute them, we replaced the steady-state activation gates previously used by their inward/outward form, defined above, in the definition of the steady-state currents. In Fig \ref{fig:7} (center), we represented the evolution of the bistability window limits ($I_1$ and $I_2$) as the maximal conductance of the current taken into account increased. The same methodology employed for the bistability window heatmaps was used to compute those limits. In Fig.~\ref{fig:7} (right), we applied hyperpolarizing steps of currents to a neuron with a fixed maximal conductance for the current considered (\ie, the inward KM current in the case of Fig.~\ref{fig:7}A (right) from $I_2$ to $I_1$, with $t_{\text{step}}=\SI{250}{\milli\second}$, starting from the resting state in $I_2$, for a total time of \SI{1}{\second}. The permeability $\bar{p}_{\mathrm{CaL}}$ used in Fig.~\ref{fig:7} was $\qty{1.5e-5}{\centi\meter\per\second}$.

\subsection*{Bifurcation analyses around the excitability switches}

    In Figs.~\ref{fig:8}E--H and Fig.~\ref{fig:9-CaLs+Kir/KM}C,D, we computed bifurcation diagrams with respect to the applied current $I$. Fixed points were computed with the \texttt{NLsolve} package for each tested $I$. Limit-cycle extrema during spiking were computed using the same methodology as for the fI curves. Conductance sets used (as $\{\bar{p}_{\mathrm{CaL}};\bar{g}_{\mathrm{KM}};\bar{g}_{\mathrm{Kir}}\}$): Fig.~\ref{fig:8}E, $\{0.;0.;0.\}$; Fig.~\ref{fig:8}F, $\{\qty{7.5e-6}{};0.;0.\}$; Fig.~\ref{fig:8}G, $\{0.;0.;0.3\}$\, ; Fig.~\ref{fig:8}H, $\{0.;0.3;0.\}$; Fig.~\ref{fig:9-CaLs+Kir/KM}C, $\{\qty{7.5e-6}{};0.;0.3\}$; Fig.~\ref{fig:9-CaLs+Kir/KM}D, $\{\qty{7.5e-6}{};0.3;0.\}$.

\section*{Acknowledgments}
Computational resources have been provided by the Consortium des Équipements de Calcul Intensif (CÉCI), funded by the Fonds de la Recherche Scientifique de Belgique (F.R.S.-FNRS) under Grant No. 2.5020.11 and by the Walloon Region. This work was supported by the Belgian Government through the Federal Public Service Policy and Support. 

\bibliography{%
    bibliography.bib%
}

\section*{Supporting information}

\begin{figure}[h!]
    \centering
    \includegraphics[width=0.5\textwidth]{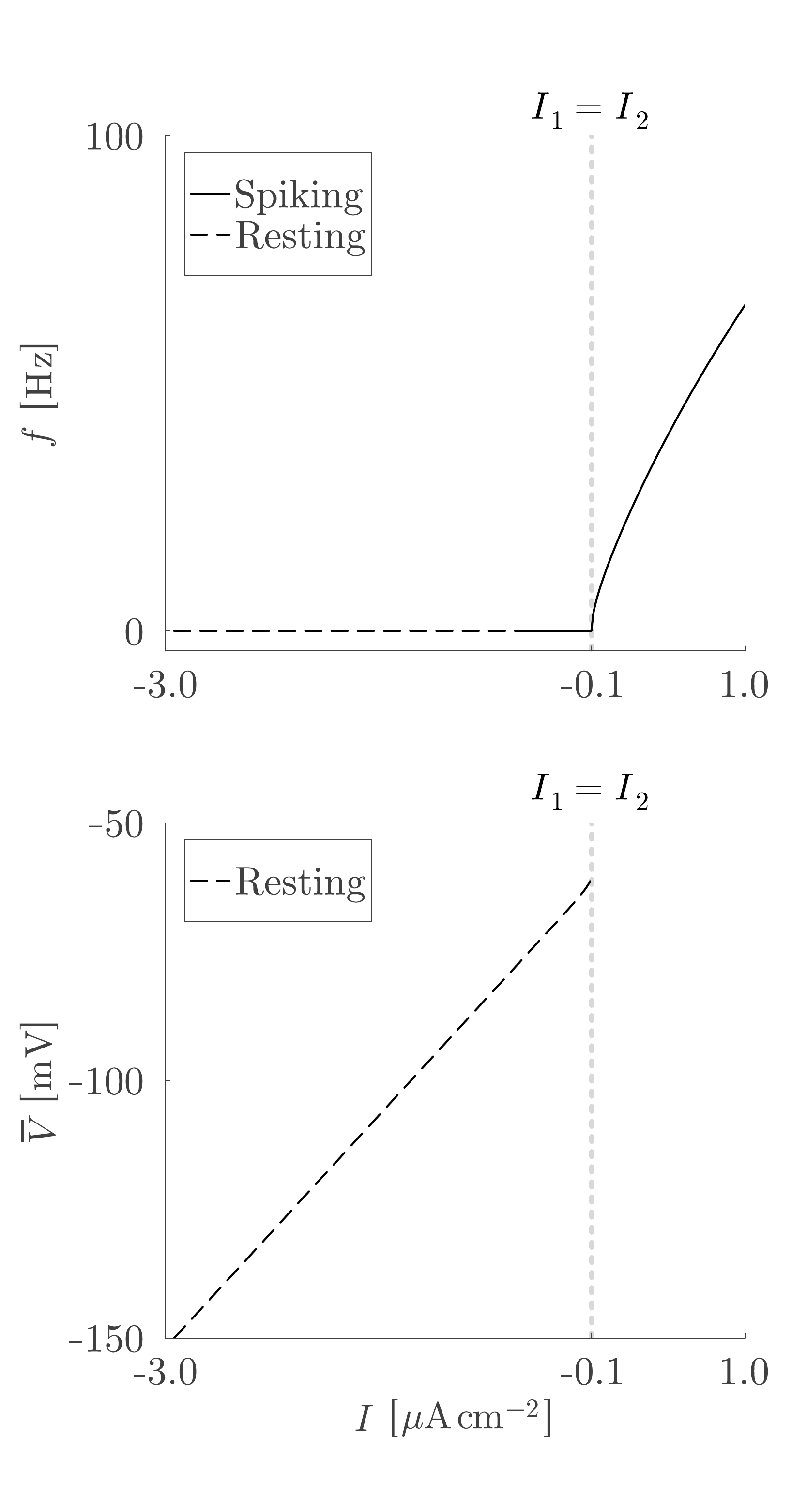}
    \caption*{\paragraph*{S1 Fig 1.}\label{app:fig:fI_hh}{{\bf Without CaL channels, the model initially did not show bitability. }The superimposition of the fI curves computed from resting and spiking, respectively, did not show the bistability window (top), that is, spiking appeared when resting disappeared as the applied current increased. The resting equilibria were progressively depolarized as the current increased (bottom).
        }}
\end{figure}

\paragraph*{S1 Appendix.}
\label{app:fI_hh}
{\bf Initially, the minimal model did not show bistability.}
    The minimal model used in this study did not exhibit bistability initially, i.e. when only voltage-gated sodium channels, delayed-rectifier potassium channels, and a leak current were considered (\nameref{app:fig:fI_hh}).

\begin{figure}[h!]
    \centering
    \includegraphics[width=\textwidth]{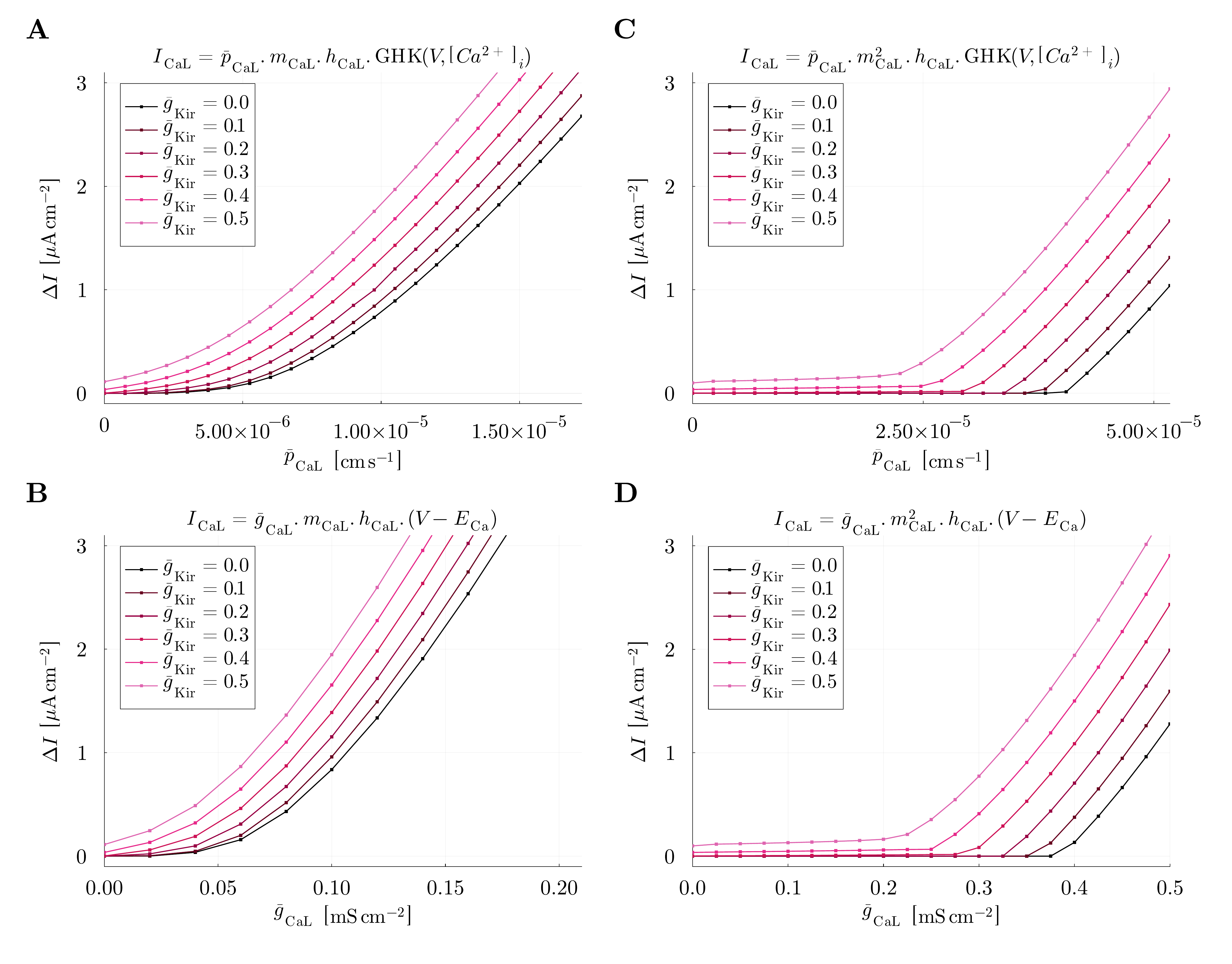}
    \caption*{\paragraph*{S2 Fig 1.} \label{app:fig:1-dI-Kir}{\bf The increase in bistability window size induced by Kir channels is observed for $q_{\mathrm{CaL}}$ of 1 and 2, and whether or not calcium-dependent interactions are considered.}
    A:~Evolution of the bistability window size $\Delta I$ for an increasing permeability $\bar{p}_{\mathrm{CaL}}$ and several values of the maximal conductance $\bar{g}_{\mathrm{Kir}}$, for a CaL current embedding a calcium-dependency and $q_{\mathrm{CaL}}=1$.
    B:~Same as A for a CaL current embedding a calcium-dependency and $q_{\mathrm{CaL}}=2$.
    C:~Same as A for a CaL current, neglecting a calcium-dependency and $q_{\mathrm{CaL}}=1$.
    D:~Same as A for a CaL current neglecting a calcium-dependency and $q_{\mathrm{CaL}}=2$.}
\end{figure}

\begin{figure}[h!]
    \centering
    \includegraphics[width=\textwidth]{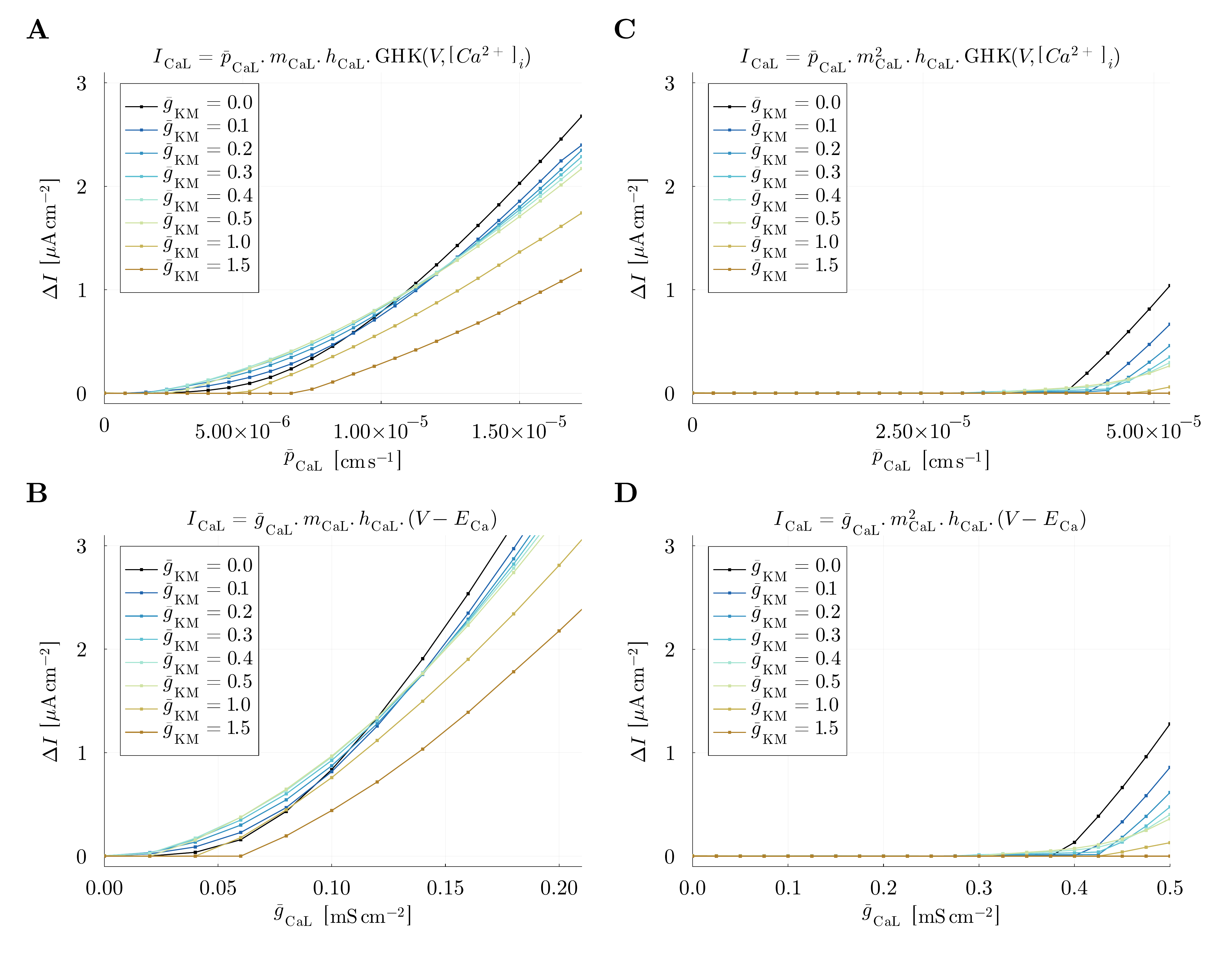}
    \caption*{\paragraph*{S2 Fig 2.}\label{app:fig:1-dI-KM}{\bf The reduction in the bistability window size induced by KM channels is observed for $q_{\mathrm{CaL}}$ of 1 and 2, and whether or not calcium-dependent interactions are considered.}
    A:~Evolution of the bistability window size $\Delta I$ for an increasing permeability $\bar{p}_{\mathrm{CaL}}$ and several values of the maximal conductance $\bar{g}_{\mathrm{KM}}$, for a CaL current embedding a calcium-dependency and $q_{\mathrm{CaL}}=1$.
    B:~Same as A for a CaL current embedding a calcium-dependency and $q_{\mathrm{CaL}}=2$.
    C:~Same as A for a CaL current neglecting a calcium-dependency and $q_{\mathrm{CaL}}=1$.
    D:~Same as A for a CaL current neglecting a calcium-dependency and $q_{\mathrm{CaL}}=2$.}
\end{figure}

\paragraph*{S2 Appendix.}
\label{app:other_CaL}
{\bf The increase and decrease of the bistability window size provided by Kir and KM channels, respectively, are still observed for several models of the CaL current.}
In \cite{le_franc_multiple_2010}, the decision to model the L-type calcium current as proportional to $m_{\mathrm{CaL}}^2$ or to $m_{\mathrm{CaL}}$ was slightly unclear. This decision was investigated here to ensure that these choices did not impact our results. 
In the same view, the impact of incorporating a Goldman–Hodgkin–Katz formalism was explored to determine whether the neglect of the calcium concentration dependency in $I_{\mathrm{CaL}}$ would alter the conclusions drawn here.
For that purpose, we modified the model of $I_{\mathrm{CaL}}$ and repeated the experiment in Fig.\ref{fig:3}A and C. The results of this experiment can be found in \nameref{app:fig:1-dI-Kir} and \nameref{app:fig:1-dI-KM}, where $GHK(V,\left[\mathrm{Ca}^{2+} \right]_i)$ denotes the Goldman–Hodgkin–Katz formalism described in the Methods section. These results showed that the increase and decrease of the bistability window size provided Kir and KM channels (respectively) in combination with CaL channels are still observed for each model of $I_{\mathrm{CaL}}$ considered.

\begin{figure}[h!]
    \centering
    \includegraphics[width=1\textwidth]{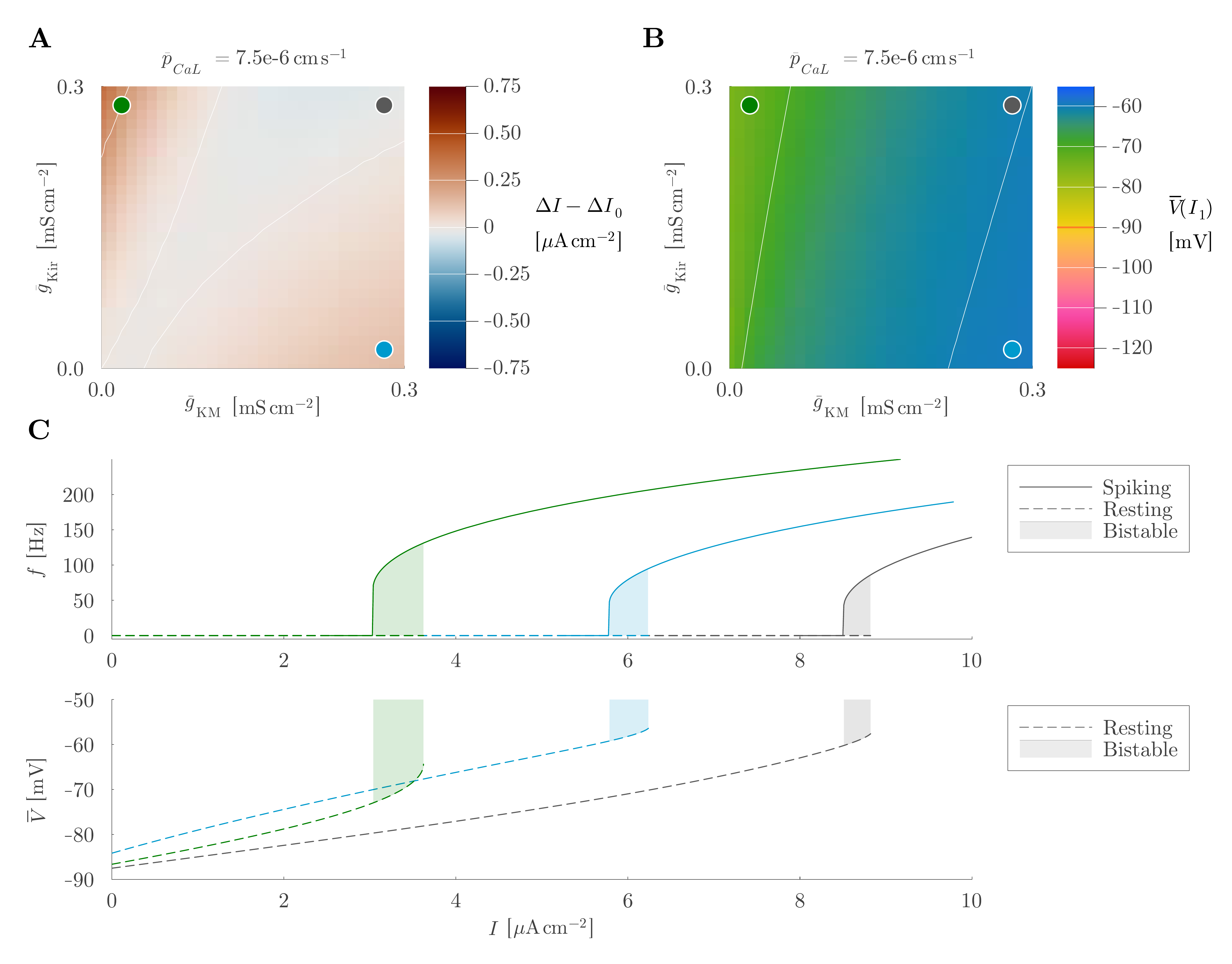} 
    \caption*{\paragraph*{S3 Fig 1.}\label{app:fig:altgt}{{\bf With an initial bistability window size of $\Delta I_0 = \qty{0.34}{\micro\ampere\per\centi\meter^2}$, Kir and KM channels increase separately the bistability window size, but promote distinct ranges of resting equilibria.} 
        A:~Heatmap of the absolute change in bistability window size ($\Delta I$) from the bistability window size resulting from $\bar{p}_\mathrm{CaL}=\qty{7.5e-6}{\centi\meter\per\second}$ and $\bar{g}_\mathrm{KM}=\bar{g}_\mathrm{Kir}=0$ (noted $\Delta I_0$, equal to $\qty{0.34}{\micro\ampere\per\centi\meter^2}$), depending on the pair of conductances $(\bar{g}_\mathrm{KM},\bar{g}_\mathrm{Kir})$ used. 
        B:~Heatmap of the resting equilibrium at the lower bound of the bistability window ($\overline{V}(I_1)$), depending on the pair of conductances $(\bar{g}_\mathrm{KM},\bar{g}_\mathrm{Kir})$ used. 
        C:~Superimposed fI curves from resting and spiking (top) and resting membrane potentials observed for a range of applied current, both displaying the bistability window, corresponding to the 3 sets of parameters marked in A and B. The white lines in A and B represent contour lines for each heatmap. 
        }}
\end{figure}

\paragraph*{S3 Appendix.}
\label{app:altgt}
{\bf Both Kir and KM channels increased the narrow bistability window resulting from a $\bar{p}_{\mathrm{CaL}}$ value that was halved compared to that used in Fig.\ref{fig:altgt}. }
    We performed a similar analysis to what was made in Fig.\ref{fig:altgt} with half the value used for $\bar{p}_{\mathrm{CaL}}$, which created a bistability window size of $\Delta I_0 = \qty{0.34}{\micro\ampere\per\centi\meter^2}$. Increasing both Kir and KM conductances between $0$ and $\qty{0.3}{\milli\siemens\per\centi\meter^2}$ showed that both Kir and KM channels may increase the bistability window size, if initially lower that what was considered n Fig.\ref{fig:altgt}. The largest increase in the bistability window size observed as $\bar{g}_\mathrm{KM}$ was increased to $\qty{0.3}{\milli\siemens\per\centi\meter^2}$ was of $\qty{0.14}{\micro\ampere\per\centi\meter}$, obtained with $\bar{g}_\mathrm{Kir}=0$ (\nameref{app:fig:altgt}A). Similar to the results shown in Fig.\ref{fig:altgt}, Kir and KM promoted different ranges of resting equilibria (\nameref{app:fig:altgt}B and C).

\begin{figure}[htbp]
    \centering
    \includegraphics[width=1\textwidth]{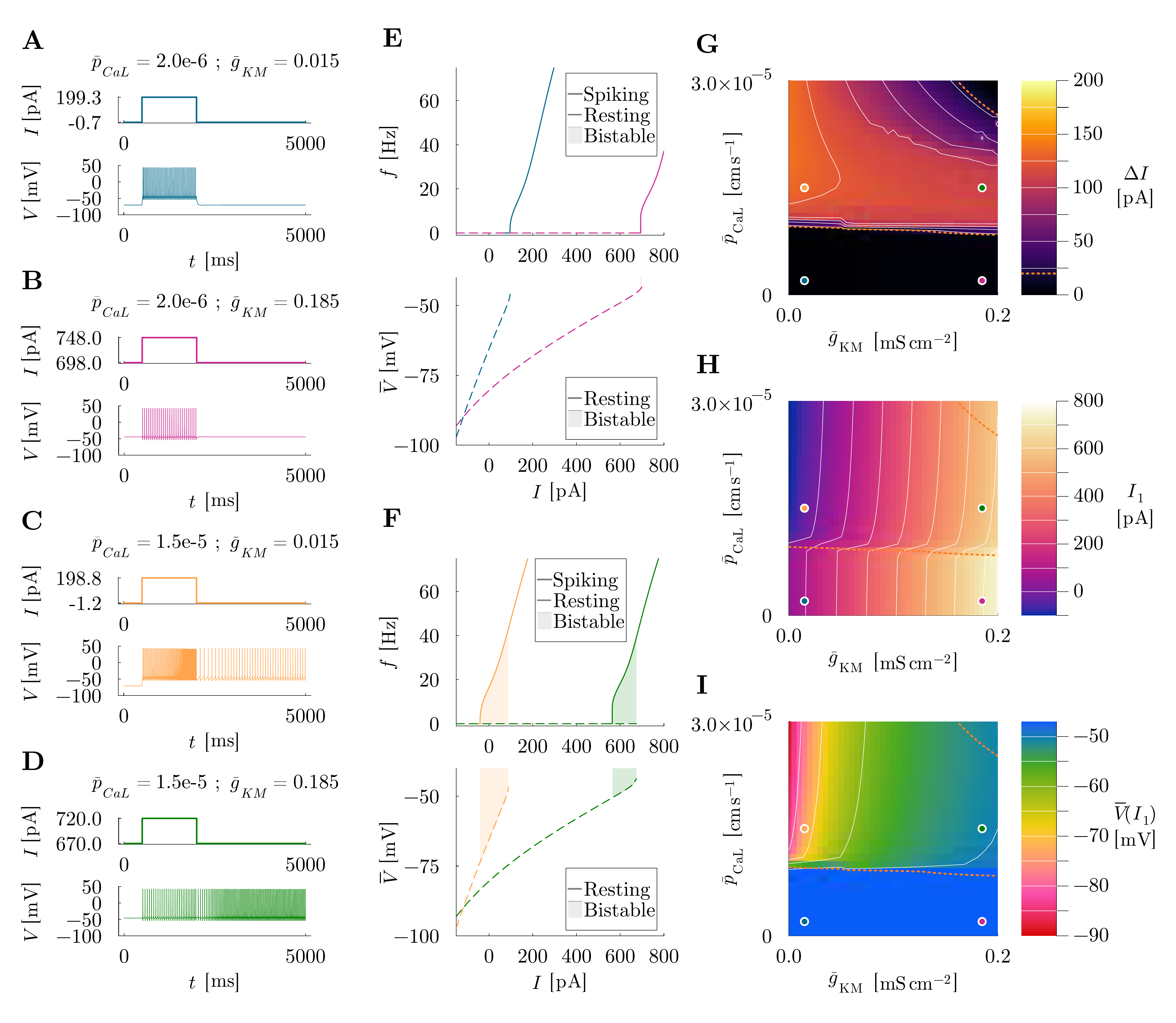} 
    \caption*{\paragraph*{S4 Fig 1.}\label{app:fig:KM_in_comp}{\bf In a more complete two-compartment model of deep projection neurons, replacing Kir by KM channels shows that KM channels slightly decrease $\Delta I$, or even decrease it to $\qty{0}{\pico\ampere}$.  } 
        A:~Response of the two-compartment model with a low $\bar{p}_\mathrm{CaL}$ and a low $\bar{g}_\mathrm{KM}$ initially at $\qty{-65}{\milli\volt}$ to a pulse of current of $\qty{200}{\pico\ampere}$. 
        B:~Same as A, but with a $\bar{g}_\mathrm{KM}$ increased to $\qty{0.185}{\milli\siemens\per\centi\meter^2}$ and a $\qty{50}{\pico\ampere}$ pulse of current. 
        C:~Same as A, but with a $\bar{p}_\mathrm{CaL}$ increased to $\qty{1.5e-5}{\centi\meter\per\second}$. 
        D:~Same as C with a $\bar{p}_\mathrm{CaL}$ increased to $\qty{1.5e-5}{\centi\meter\per\second}$. 
        E:~Superimposed fI curves from resting and spiking (top) and resting membrane potentials observed for a range of applied current (bottom), both displaying the bistability window (if existing), corresponding to the 2 sets of parameters used in A and C. 
        F:~Same as E, but for the 2 sets of parameters used in B and D. 
        G:~Heatmap of the bistability window size ($\Delta I$) depending on the pair of conductances $(\bar{g}_\mathrm{KM},\bar{p}_\mathrm{CaL})$ used. 
        H:~Heatmap of the bistability window lower limit ($I_1$) obtained for each pair of conductances. 
        I:~Heatmap of the resting equilibrium observed at bistability window lower limit ($\overline{V}(I_1)$) for each pair of conductances.
        Each heatmap contained $41 \times 31$ data points. The white lines in G, H and I represent contour lines for each heatmap. The red dashed line marks $\Delta I =\qty{20}{\pico\ampere}$. } 
\end{figure}

\paragraph*{S4 Appendix.}
\label{app:KM_in_comp}
{\bf Replacing Kir by KM channels in the complete conductance-based model shows that KM channels decrease the bistability window size.} We performed the same experiments as in Fig.~\ref{fig:comp} but replaced the Kir channels with KM channels. For low $\bar{p}_{\mathrm{CaL}}$, there was no bistability observed in the pulse responses (\nameref{app:fig:KM_in_comp}A and B). Increasing $\bar{p}_{\mathrm{CaL}}$ allowed to observed bistability, for lower and higher values of $\bar{g}_{\mathrm{KM}}$ (\nameref{app:fig:KM_in_comp}C and D). The superimposed fI curves confirmed the absence of bistability for low $\bar{p}_{\mathrm{CaL}}$, for the two values of $\bar{g}_{\mathrm{KM}}$ considered (\nameref{app:fig:KM_in_comp}E, top). Increasing $\bar{p}_{\mathrm{CaL}}$ resulted in the superimposed fI curves displaying a bistability window for the two values of $\bar{g}_{\mathrm{KM}}$ (\nameref{app:fig:KM_in_comp}F, top). Increasing $\bar{g}_{\mathrm{KM}}$ starting from either of the values of $\bar{p}_{\mathrm{CaL}}$ used increased the resting equilibria observed (\nameref{app:fig:KM_in_comp}E and F, bottom). Using a range of $\bar{p}_{\mathrm{CaL}}$ and $\bar{g}_{\mathrm{KM}}$ showed that KM channels decrease slightly the size of the bistability window for $\bar{p}_{\mathrm{CaL}} \approx \qty{1.5e-5}{\centi\meter\per\second}$ (\nameref{app:fig:KM_in_comp}G), which is similar to the results obtained with the minimal model in Fig.\ref{fig:3}A. For $\bar{p}_{\mathrm{CaL}}$ above $\qty{2e-5}{\centi\meter\per\second}$, the reduction is $\Delta I$ is stronger as $\bar{g}_{\mathrm{KM}}$ increases. Whereas CaL channels showed a monotonic increase in $\Delta I$ for $\bar{g}_{\mathrm{KM}}=0$, increasing $\bar{g}_{\mathrm{KM}}$ broke this monotonicity. In fact, the increase of $\Delta I$ for $\bar{p}_{\mathrm{CaL}} \approx \qty{1.5e-5}{\centi\meter\per\second}$ dropped as $\bar{p}_{\mathrm{CaL}}$ was increased further. KM channels also increased $I_1$ (
\nameref{app:fig:KM_in_comp}H) and $\overline{V}(I_1)$ (\nameref{app:fig:KM_in_comp}I).

\begin{figure}[htbp]
    \centering
    \includegraphics[width=1\textwidth]{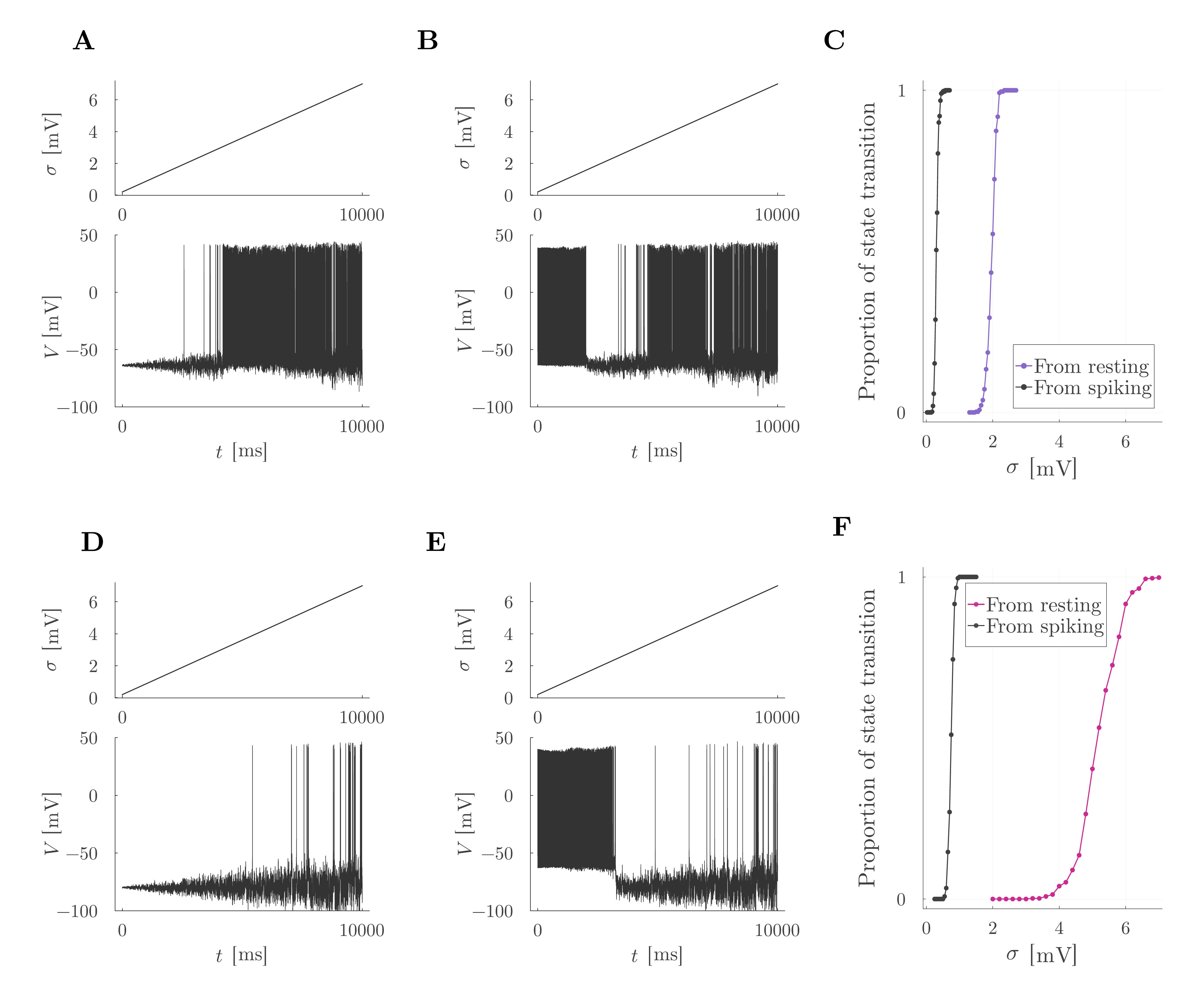} 
    \caption*{\paragraph*{S5 Fig 1.}\label{app:fig:noise_I1}{\bf 
    At the beginning of the bistability window, resting states are more robust to noise than spiking states considering whether KM or Kir channels are considered in combination with CaL channels.} 
        A:~Response of the model including $I_\mathrm{KM}$ to $I_1$ with an additional white noise of increasing standard deviation $\sigma$, when initially resting. B:~Same as A, when initially spiking. C:~Proportion of trials that failed to maintain their initial behavior (resting or spiking) and transitioned to spiking or resting, respectively, as a function of the value of constant standard deviation $\sigma$ used to create white noise, with the model including $I_\mathrm{KM}$. D:~Same as A, but with the model including $I_\mathrm{Kir}$ instead of $I_\mathrm{KM}$. E:~Same as A, but with the model including $I_\mathrm{Kir}$ instead of $I_\mathrm{KM}$, and when initially spiking. F:~Same as C, but with the model including $I_\mathrm{Kir}$ instead of $I_\mathrm{KM}$. 
        The bistability window size in A, B, and C, is similar to those of D, E, and F (resulting from the addition of $I_\mathrm{Kir}$), such that the same value of maximal conductance was used for $\bar{g}_\mathrm{KM}$ (in A, B, and C) and for $\bar{g}_\mathrm{Kir}$ (in D, E, and F), but with a higher $\bar{p}_\mathrm{CaL}$ in A, B, and C compared to D, E, and F.}
\end{figure}

\paragraph*{S5 Appendix.}
\label{app:noise_I1}
{\bf Resting states are more robust to noise than spiking states at $I_1$ considering whether KM or Kir channels are considered in combination with CaL channels.} Here, we performed the same experiments as in Fig.~\ref{fig:4} but with $I_1$ instead of the central current of the bistability window (\ie $\frac{I_1+I_2}{2}$) considering either KM or Kir channels in combination with CaL channels (\nameref{app:fig:noise_I1}). This experiment quantifies the robustness to noise of resting and spiking for $\bar{g}_\mathrm{KM}$-bistability and $\bar{g}_\mathrm{Kir}$-bistability. Compared to Fig.~\ref{fig:4}, the resting states were more robust at $I_1$ than at $\frac{I_1+I_2}{2}$, but the spiking states were more robust at $\frac{I_1+I_2}{2}$ than at $I_1$, with both types of bistability. This is consistent with a higher(resp. lower) reachability of the resting(resp. spiking) state at $I_1$ than at $\frac{I_1+I_2}{2}$ as observed in Fig.\ref{fig:6B}. For both $\bar{g}_\mathrm{KM}$-bistability and for $\bar{g}_\mathrm{Kir}$-bistability, the resting state in $I_1$ was more robust than the spiking state. This may be due to the higher reachability of the resting state than of the spiking state in $I_1$, as observed in Fig.\ref{fig:6B}. Finally, these results showed that for both in $I_1$ and in $\frac{I_1+I_2}{2}$, resting with $\bar{g}_\mathrm{Kir}$-bistability was more robust than resting with $\bar{g}_\mathrm{KM}$-bistability, consistently with a higher reachability throughout the bistability window of the resting state with $\bar{g}_\mathrm{Kir}$-bistability compared to that with $\bar{g}_\mathrm{KM}$-bistability.

\paragraph*{S6 Appendix.}
\label{app:CaLf}
{\bf The other faster L-type calcium current described by \cite{le_franc_multiple_2010} has a reduced impact on bistability.}
    \cite{le_franc_multiple_2010} described and modeled two types of L-type calcium channels. In this study, only the slower type of L-type calcium channels was considered, but whether the faster type was also able to create bistability was unclear. To explore its effect on bistability, we added the ionic current arising from the faster type of L-type calcium channels ($I_{\mathrm{CaL,f}}$) to the model. The fast L-type calcium current, embedding a calcium-dependency, was modeled using a Goldman–Hodgkin–Katz formalism \cite{hodgkin_effect_1949} such that: 
    \begin{linenomath}
    \begin{eqnarray}
        I_{\mathrm{CaL,f}} = \bar{p}_\mathrm{CaL,f} \, m_{\mathrm{CaL,f}}^{2} \, h_{\mathrm{CaL,f}} \mathrm{GHK}(V,\left[\mathrm{Ca}^{2+} \right]_i) ,
    \end{eqnarray}
    \end{linenomath}
    \begin{linenomath}
    \begin{eqnarray}
        \text{with }\mathrm{GHK}(V,\left[\mathrm{Ca}^{2+} \right]_i) =  10^{-3} \cdot 2F \cdot \left( \left[\mathrm{Ca}^{2+} \right]_i \cdot \dfrac{-w}{e^{-w} -1} - \left[\mathrm{Ca}^{2+} \right]_o \cdot \dfrac{w}{e^{w} -1}\right) ,
    \end{eqnarray}
    \end{linenomath}
    \begin{linenomath}
    \begin{eqnarray}
        \text{and } w = 10^{-3}\cdot V\cdot \dfrac{2F}{RT}.
    \end{eqnarray}
    \end{linenomath}
    Each variable had the same value and meaning as described in the Methods section. The steady-state curves of the activation $m_{\mathrm{CaL,f}}$ and inactivation $h_{\mathrm{CaL,f}}$ can be found in \nameref{app:tab:minfandtau-CaLf}. The evolution of the intracellular calcium concentration was modified to take $I_{\mathrm{CaL,f}}$ into account as part of the total calcium current flowing through the membrane, such that:
    \begin{linenomath}
    \begin{eqnarray}
        \dfrac{d\left[\mathrm{Ca}^{2+} \right]_i}{dt} = - 10^7 \cdot \dfrac{\left(I_{\mathrm{CaL}}+I_{\mathrm{CaL,f}}\right)}{10^3} \cdot \dfrac{1}{2Fd} + \dfrac{\left(\left[\mathrm{Ca}^{2+} \right]_{i,0}-\left[\mathrm{Ca}^{2+} \right]_i \right) }{\tau_\mathrm{Ca}}.
    \end{eqnarray}
    \end{linenomath}

\begin{figure}[htbp]
    \centering
    \includegraphics[width=\textwidth]{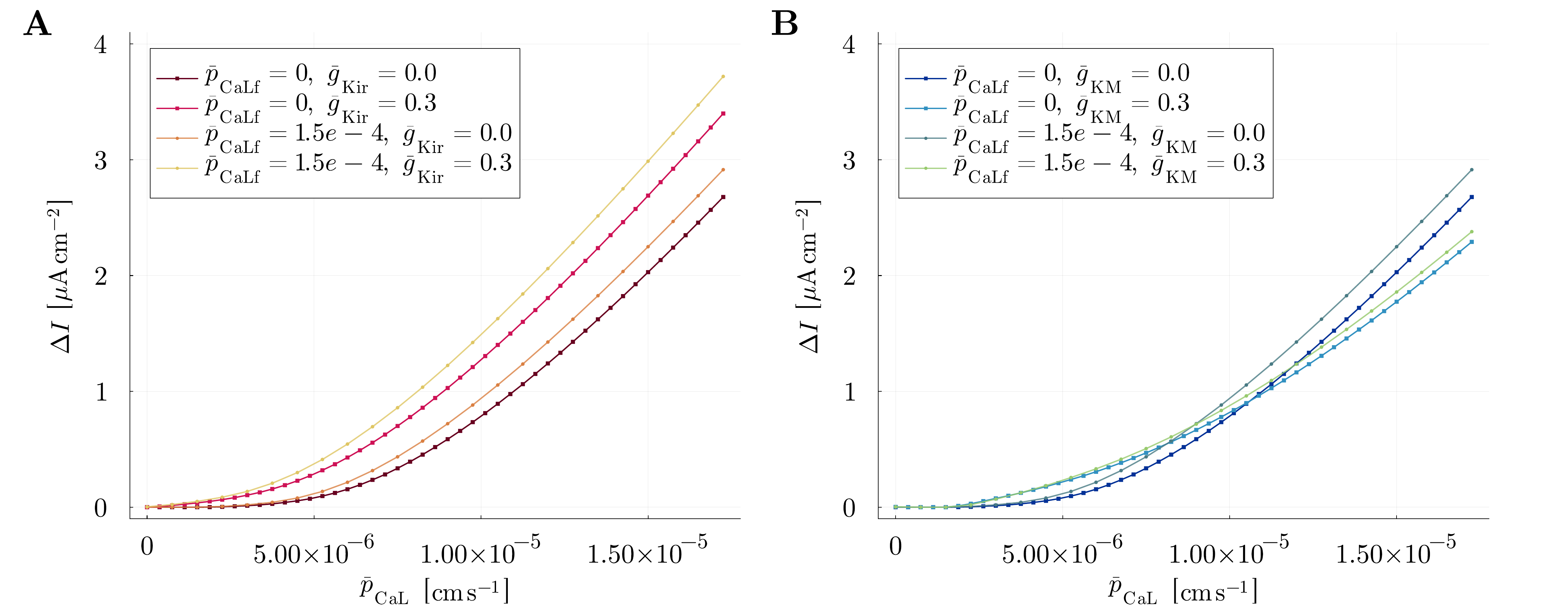}
    \caption*{\paragraph*{S6 Fig 1.} \label{app:fig:2-pCaLf}{\bf The other L-type calcium current ($I_{\mathrm{CaL,f}}$) described by \cite{le_franc_multiple_2010} has a reduced impact on bistability compared to the L-type calcium current considered in this work ($I_{\mathrm{CaL,s}}$), whether or not the Kir or KM current were considered.}
    A:~Evolution of the bistability window size $\Delta I$ for an increasing permeability $\bar{p}_{\mathrm{CaL,s}}$, two values of the maximal conductance $\bar{g}_{\mathrm{Kir}}$, and two values of permeability $\bar{p}_{\mathrm{CaL,f}}$ of the other type of L-type calcium channels.
    B:~Same as A with several values of $\bar{g}_{\mathrm{KM}}$ instead of $\bar{g}_{\mathrm{Kir}}$.}
\end{figure}

    To assess the effect of $I_{\mathrm{CaL,f}}$ on bistability, the experiment realized in Fig.\ref{fig:3} A and C was repeated with two fixed value of permeability $\bar{p}_{\mathrm{CaL,f}}$ or \qty{0}{} and \qty{1.5e-4}{\centi\meter\per\second}. As observed in \nameref{app:fig:2-pCaLf}, the faster L-type calcium current did not create bistability itself; rather, it appeared to widen the bistability window engendered by the slower L-type calcium current. 

    \begin{table}[!ht]
        \begin{adjustwidth}{-2.25in}{0in} 
        \centering
        \caption*{\paragraph*{S6 Table 1.} \label{app:tab:minfandtau-CaLf}
        {\bf Steady-state activations and inactivation used to model $I_{\text{CaLf}}$.}}
        \begin{tabular}{|c|c|c|}
        \hline
        $x_{\mathrm{ion}}$ & $x_{\mathrm{ion},\infty}(V)$ & $\tau_{x_{\mathrm{ion}}}(V)$ \\ \thickhline
        $m_{\text{CaLf}}$ & $\dfrac{1}{1 + \exp(-(V+\num{17.5})/4.3)}$ & $\dfrac{0.5}{\frac{0.1(-V-40)}{\exp{(0.1(-V-40))} -1 + 4 \exp{((-V-65)/18)}}}$ \\ \hline
        $h_{\text{CaLf}}$ & $\dfrac{1}{1 + \exp((V+14)/\num{4.03})}$ & $1500$ \\ \hline
        \end{tabular}
        \begin{flushleft} 
        \end{flushleft}
        \end{adjustwidth}
    \end{table}

\begin{figure}[htbp]
    \centering
    \includegraphics[width=0.5\textwidth]{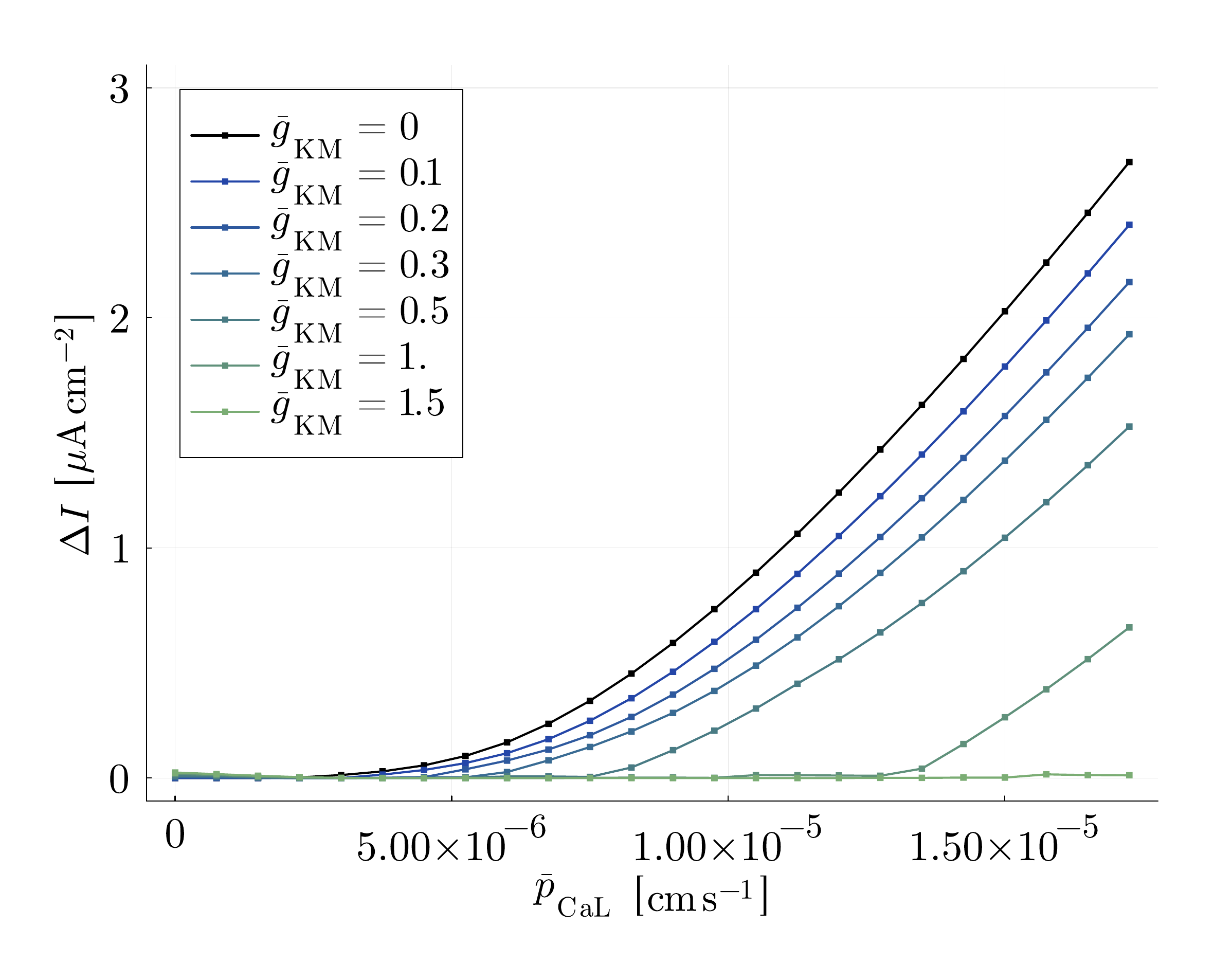}
    \caption*{\paragraph*{S7 Fig 1.}\label{app:fig:3-mKM}{\bf The reduction in the bistability window size induced by KM channels is robust to a change in the KM channel steady-state activation to $m_{\mathrm{KM,\infty}}= \frac{1}{1+\exp{((-V-26.7)/12.6)}}$.} Evolution of the bistability window size $\Delta I$ for an increasing permeability $\bar{p}_{\mathrm{CaL}}$, and several values of the maximal conductance $\bar{g}_{\mathrm{KM}}$.}
\end{figure}

\paragraph*{S7 Appendix.}
\label{app:other_KM}
{\bf The reduction in the bistability window size is still provided by KM channels opening at a higher membrane potential.}
    In this study, the steady-state activation curve of KM channels was modified to allow the comparison with Kir channels within the same range of potentials. However, to demonstrate that the decrease in the bistability window size caused by the addition of KM channels was still present without this modification, the experiment realized for Fig.\ref{fig:3}A and C was repeated with the unmodified steady-state activation curve of KM channels, which can be found in \nameref{app:tab:minfandtau-KM}. Those results can be found in \nameref{app:fig:3-mKM}.

    \begin{table}[!ht]
        \begin{adjustwidth}{-2.25in}{0in} 
        \centering
        \caption*{\paragraph*{S7 Table 1.} \label{app:tab:minfandtau-KM}
        {\bf Steady-state activation used to model $I_{\text{KM'}}$.}}
        \begin{tabular}{|c|c|c|}
        \hline
        $x_{\mathrm{ion}}$ & $x_{\mathrm{ion},\infty}(V)$ & $\tau_{x_{\mathrm{ion}}}(V)$ \\ \thickhline
        $m_{\text{KM'}}$ & $\dfrac{1}{1 + \exp((-V-\num{26,7})/\num{12,6})}$ & $103$ \\ \hline
        \end{tabular}
        \begin{flushleft} 
        \end{flushleft}        
        \end{adjustwidth}
    \end{table}

\end{document}